\documentclass[twocolumn]{autart}    

\usepackage{graphicx}          

\usepackage{amssymb,amsmath,graphicx,multicol,epstopdf}

\newtheorem{Example}{Example}[section]
\newtheorem{Remark}{Remark}[section]
\newtheorem{Corollary}{Corollary}[section]
\newtheorem{Definition}{Definition}[section]
\newtheorem{Theorem}{Theorem}[section]
\newtheorem{Proposition}{Proposition}[section]
\newtheorem{proof}{Proof}
\newtheorem{Lemma}{Lemma}[section]
\newtheorem{Assumption}{Assumption}[section]

\newcommand{\N}{{\mathcal N}}

\newcommand{\BM}{\begin{matrix}}
\newcommand{\EM}{\end{matrix}}

\newcommand{\ba}{\begin{array}}
\newcommand{\ea}{\end{array}}
\newcommand{\be}{\begin{eqnarray}}
\newcommand{\ee}{\end{eqnarray}}
\newcommand{\EQQ}{\begin{eqnarray*}}
\newcommand{\ENN}{\end{eqnarray*}}

\newcommand{\R}{{\mathbb R}}

\DeclareMathOperator*\argmin{arg\,min}

\newcommand{\rank}[1]{rank [{#1}]}

\def\tt#1#2{\left[\begin{array}{c}#1\\ #2\end{array}\right]}

\def\rank{\mbox{rank}}

\newcommand{\mean}[1]{\mathbb{E}(#1)}
\newcommand{\var}[1]{\mathbb{V}(#1)}
\newcommand{\cov}[1]{\mathbb{C}ov(#1)}

\newcommand{\bd}{\begin{Definition}
\begin{rm} }
\newcommand{\ed}{ \end{rm}
\end{Definition} }

\newcommand{\bexercise}{\begin{exercise}\vspace{-3mm}
\begin{rm} }
\newcommand{\eexercise}{ \end{rm}
\end{exercise} }

\newcommand{\bappend}{\begin{append}\vspace{-3mm}
\begin{rm} }
\newcommand{\eappend}{ \end{rm}
\end{append} }

\newcommand{\basm}{\begin{Assumption} \begin{rm}}
\newcommand{\easm}{\end{rm} \end{Assumption}}

\newcommand{\bpropty}{\begin{property} \vspace{-3mm}\begin{rm}}
\newcommand{\epropty}{\end{rm} \end{property}}

\newcommand{\bremark}{\begin{Remark}
\begin{rm} }
\newcommand{\eremark}{\hfill$\lozenge$ \end{rm}\end{Remark} }
\newcommand{\bt}{\begin{Theorem} \begin{rm} }
\newcommand{\et}{ \end{rm}
\end{Theorem} }
\newcommand{\bl}{\begin{Lemma} \begin{rm} }
\newcommand{\el}{ \end{rm}
\end{Lemma} }
\newcommand{\bcorollary}{\begin{Corollary} \begin{rm} }
\newcommand{\ecorollary}{ \end{rm}
\end{Corollary} }
\newcommand{\bdefinition}{\begin{Definition}\begin{rm} }
\newcommand{\edefinition}{ \end{rm}
\end{Definition} }
\newcommand{\bproposition}{\begin{proposition} \begin{rm} }
\newcommand{\eproposition}{ \end{rm}
\end{proposition} }
\newcommand{\bexample}{\begin{example} \begin{rm} }
\newcommand{\eexample}{ \end{rm}
\end{example} }

\newcommand{\bproof}{\begin{proof} \begin{rm} }
\newcommand{\eproof}{ \end{rm} \end{proof} }

\DeclareMathOperator*{\minimize}{minimize}
\DeclareMathOperator*{\maximize}{maximize}

\begin{document}
\begin{frontmatter}

\title{On kernel design for regularized LTI system identification\thanksref{footnoteinfo}}

\thanks[footnoteinfo]{Abridged and preliminary results of this paper was presented in the 19th
IFAC World Congress, Cape Town, South Africa, 2014, the European Research Network on System Identification (ERNSI) workshop, Ostend, Belgium,
2014, and the 17th IFAC Symposium on System Identification, Beijing, China, 2015.
\\
This work is supported by the Thousand Youth Talents Plan funded by
the central government of China, the Shenzhen Projects Ji-20170189 and Ji-20160207 funded by Shenzhen Science and Technology Innovation Council, 
the President's grant under contract No. PF. 01.000249 and the
Start-up grant under contract No. 2014.0003.23 funded by the Chinese
University of Hong Kong, Shenzhen, as well as by a research grant
for junior researchers under contract No. 2014-5894, funded by
Swedish Research Council.}

\author{Tianshi Chen} 

\address{School of Science and Engineering, The Chinese University of Hong Kong, Shenzhen, 518172, China, (e-mail:
tschen@cuhk.edu.cn).}


\begin{abstract}                
There are two key issues for the kernel-based regularization method:
one is how to design a suitable kernel to embed in the kernel the
prior knowledge of the LTI system to be identified, and the other
one is how to tune the kernel such that the resulting regularized
impulse response estimator can achieve a good bias-variance
tradeoff. In this paper, we focus on the issue of kernel design.
Depending on the type of the prior knowledge, we propose two methods
to design kernels: one is from a machine learning perspective and
the other one is from a system theory perspective. We also provide
analysis results for both methods, which not only enhances our
understanding for the existing kernels but also directs the design
of new kernels.
\end{abstract}


\end{frontmatter}


\section{Introduction}


Among diverse system identification problems, the linear
time-invariant (LTI) system identification is a classical and
fundamental problem. As well-known, an LTI system is uniquely
characterized by its impulse response, and thus the LTI system
identification is equivalent to its impulse response estimation that
could be ill-conditioned in practice. So to tackle the LTI system
identification, one needs to first find a way to make the
ill-conditioned problem well-conditioned. Then one faces the core
issue of system identification, i.e., how to construct a model
estimator able to achieve a good bias-variance tradeoff, or
equivalently, to suitably balance the adherence to the data and the
model complexity.

There are different routes to handle these two issues. The most
widely used route is to adopt instead a parametric model structure
with a finite number of parameters. The so-called maximum
likelihood/prediction error method (ML/PEM) is one method along this
route. It has optimal asymptotic properties and is the current
standard method to LTI system identification. It first postulates a
parametric model structure, e.g., a rational transfer function, then
forms the prediction error criterion, and finally obtains the model
estimator by minimizing the prediction error criterion; see e.g.,
\cite{Ljung:99,SoderstromS:89}. The model complexity of the
parametric model structure is governed by the number of parameters
and tuned in a discrete manner. The bias-variance tradeoff issue is
handled by cross validation or model structure selection criteria
such as Akaike's information criterion (AIC), Bayesian information
criterion (BIC), and etc, which correspond to combinatorial
optimization problems; see e.g., \cite[Chapter 16]{Ljung:99}.
However, these techniques are not as reliable as expected when the
data is short and/or has low signal-to-noise ratio; see e.g.,
\cite{COL12a,PDCDL14}.

Another route is to adopt regularization by adding an extra
regularization term in the estimation criterion. The kernel-based
regularization method (KRM) proposed in the seminal paper
\cite{PN10a} and further developed in \cite{PCN11,COL12a,CALCP14} is
one method along this route. It first proposes a kernel
parameterized by some hyper-parameters for the impulse response,
then estimates these hyper-parameters with certain methods, and
finally obtains the estimate of the impulse response by minimizing a
regularized least squares criterion with the regularization term in
quadratic form and the regularization matrix defined through the
kernel; see \cite{PDCDL14} for a recent survey. While the impulse
response model is used, the underlying model structure is determined
by the kernel. The model complexity is governed by the
hyper-parameters of the kernel, and tuned in a continuous manner by
e.g., the empirical Bayes method, Stein's unbiased risk estimator
\cite{PDCDL14,PC16,MCL16} and etc., which correspond to non-convex
but non-combinatorial optimization problems.

The route to adopt regularization is by no means new, see e.g.,
\cite{TichonovA:77}, \cite{SML93}, \cite[p. 504-505]{Ljung:99} and
also \cite{Chiuso16} for a historic review, but no important
progress along this route has been reported until \cite{PN10a}. The
major obstacle is that it was unclear whether or not it is possible
to design the regularization to embed the prior knowledge of the LTI
system to be identified. The intriguing finding disclosed by the KRM
is that when considering impulse response estimation problem, it is
possible to design a regularization in quadratic form or
equivalently a kernel to embed the prior knowledge of the
\emph{impulse response} of the LTI system to be identified. So far
several kernels have been invented to embed various prior knowledge,
e.g.,
\cite{PN10a,PCN11,COL12a,CALCP14,Dinuzzo12,CL14,CL15a,CL15b,CACCLP16,CCL17,MSS16}
and some analysis results of the corresponding kernels have been
derived, e.g.,
\cite{PN10a,COL12a,CALCP14,Dinuzzo12,CL14,CL15a,CL15b,CACCLP16,CCL17}.
Still, there lack systematic ways to design kernels to embed in the
kernel the prior knowledge of the impulse response of the LTI system
to be identified.

In this paper, we try to develop systematic ways to design kernels.
Clearly, this issue relates to the type of the prior knowledge.
Interestingly, even for the same impulse response estimation
problem, users with different background may come up with different
types of prior knowledge because they may treat the impulse response
in different ways. For instance, users from machine learning may
treat the impulse response as a function, whose amplitude varies
with a certain rate and decays exponentially to zero in the end;
users from signals and systems may associate the impulse response
with an LTI system that is stable and may be overdamped,
underdamped, has multiple distinct time constants and resonant
frequencies, and etc. Accordingly, we provide two methods to design
kernels to embed the corresponding type of prior knowledge from a
machine learning perspective and from a system theory perspective,
respectively.

To develop the machine learning method, we first point out a common
feature of the two most widely used single kernels, i.e., the stable
spline (SS) kernel in \cite{PN10a} and the diagonal-correlated (DC)
kernel in \cite{COL12a}, that they belong to the class of the
so-called amplitude modulated locally stationary (AMLS) kernel. The
AMLS kernel is a multiplication of two kernels: a rank-1 kernel and
a stationary kernel. It has a neat implication: the zero mean
Gaussian process with the AMLS kernel as the covariance function can
be seen as an amplitude modulated (by the rank-1 kernel) zero mean
Gaussian processes with the stationary kernel as covariance
function; see Section \ref{sec:ml_perspective} for details. This
finding leads to the machine learning method to design general
kernels: design the rank-1 kernel and the stationary kernel to
account for the decay and the varying rate of the impulse response,
respectively.


To develop the system theory method, we recall the general guideline
to design a kernel in \cite{COL12a} that is to let the kernel mimic
the behavior of the optimal kernel \cite[Thm. 1]{COL12a} and
\cite[Prop. 19]{PDCDL14}. Moreover, the prior knowledge for the
impulse response, or equivalently, the LTI system to be identified,
shall be made use of, since the optimal kernel depends on the true
impulse response. Following this guideline and employing the
multiplicative uncertainty configuration in robust control, see e.g.
\cite{ZhouDG:96}, we design the simulation induced (SI) kernel. In
particular, the prior knowledge is embedded in the nominal model,
the uncertainty is assumed to be stable and finally, the
multiplicative uncertainty configuration is used to take into
account both the nominal model and the uncertainty and is simulated
with an impulsive input to get the SI kernel. Then we make analysis
for SI kernels in terms of stability, maximum entropy property, and
Markov property, with the SS and DC kernels as examples.  In
particular, we give conditions under which the SI kernel is stable,
solves a suitably defined entropy maximization problem, and induces
a Gaussian Markov process. The maximum entropy interpretation
enhances our understanding about a kernel, and the Markov property
of a kernel ensures that the inverse of the kernel matrix is banded
and this special structure can be exploited to derive more stable
and efficient implementation for the KRM, see e.g.,
\cite{CL:12,CCL17}.

The remaining part of this paper is organized as follows. In Section
\ref{sec:krm_recap}, the KRM is recapped. The machine learning and
system theory methods to design kernels are introduced in Section
\ref{sec:ml_perspective} and \ref{sec:st_perspective}, respectively.
Then a case study is provided in Section \ref{sec:casestudy} to
demonstrate how to design kernels from the proposed two methods to
model impulse responses with damped oscillation. Finally, this paper
is concluded in Section \ref{sec:conclusion} with a take-home
message. The proofs of all propositions and theorems are given in
the Appendix.

\section{LTI System Identification with Kernel-based Regularization Method}\label{sec:krm_recap}

\subsection{Impulse Response Estimation}
Consider a single-input-single-output, causal and LTI system described by
\begin{align}\label{eq:sys}
y(t) &= (g\ast u)(t)+ v(t), \quad t=T_s,2T_s,\cdots,NT_s,
\end{align}
where $t$ is the time index, $y(t),u(t),v(t)\in\R$ and $g(t)\in\R$ are
the measured output, input, disturbance, and impulse response of the LTI system at time $t$, respectively, $(g\ast u)(t)$ is the
convolution between the impulse response $g(\cdot)$ and the input
$u(\cdot)$ evaluated at time $t$, and $T_s$ is the sampling period
and for convenience chosen to be 1.

The system (\ref{eq:sys}) represents both discrete-time (DT) system with $t=1,2,\cdots,$ and continuous-time system (CT) with
$t\geq0$. In particular, we have \begin{align}\label{eq:conv}
  (g\ast u)(t) = \left\{
                   \begin{array}{cc}
                     \sum_{\tau=1}^\infty g(\tau)u(t-\tau), & \text{ DT}, \\
                     \int_0^\infty g(\tau)u(t-\tau)d\tau, & \text{ CT}, \\
                   \end{array}
                 \right.
\end{align}  where the unmeasured portions of $u(t)$ are set to zero.
Moreover, the system (\ref{eq:sys}) is assumed to be stable, i.e., $g\in \ell^1$ for the DT case and
$g\in\mathcal L^1$  for the CT case, where $\ell^1$ denotes the space of absolutely convergent sequences and $\mathcal L^1$ denotes the space of
absolutely Lebesgue integrable functions on $t\geq0$, and the disturbance $v(t)$ is assumed to be a white Gaussian noise with zero mean and variance $\sigma^2$ and independent of $u(t)$.

The goal of LTI system identification is to estimate the impulse response $g(t)$ as well as possible given the data
$y(t),u(t)$ with $t=1,2,\cdots,N$ for the DT case and $y(t)$ with $t=1,2,\cdots,N$ and $u(t)$ with $t\geq0$ for the CT case.

\subsection{Kernel-based Regularization Method}

As well-known \cite{TichonovA:77}, \cite[p. 504-505]{Ljung:99}, straightforward impulse response estimation, i.e.,
\begin{align*}\minimize_{g} \sum_{t=1}^N (y(t) - (g*u)(t))^2
\end{align*}
could be an ill-conditioned problem in practice and one way to overcome this problem is to adopt regularization. Moreover, the recent progresses for KRM \cite{PN10a,PCN11,COL12a,CALCP14} show that if the regularization is well designed and tuned, the resulting regularized impulse response estimator can also achieve a good bias-variance tradeoff.

To introduce the KRM, we first recall the definition of the positive semidefinite kernel and
its associated reproducing kernel Hilbert space (RKHS). Let $(X, d)$
be a metric space with $d$ being its metric. A function $k:X\times
X\rightarrow \R$ is called a positive semidefinite kernel, if it is
symmetric and satisfies $\sum_{i,j=1}^m a_ia_j
k(x_i,x_j)\geq0$ for any finite set of points
$\{x_1,\cdots,x_m\}\subset X$ and $\{a_1, ..., a_m\} \subset \R$. As
well-known from e.g., \cite{Aronszajn50}, to every positive
semidefinite kernel $k$ there corresponds to one and only one class
of functions with a unique determined inner product in it, leading
to the so-called reproducing kernel Hilbert space (RKHS) $\mathcal
H_k$ with $k$ as the reproducing kernel.

The KRM first introduces a suitable positive semidefinite kernel $k(t,s;\theta)$\footnote{Sometimes the dependence of $k(t,s;\theta)$ on $\theta$ is ignored.}with $t,s\in X$, where for the DT case, $X=\mathbb N$ and for the CT case, $X=\{t|t\geq0\}$,
and $\theta\in\R^m$ is a hyper-parameter vector that contains the parameters used to parameterize the kernel, and then solves the following regularized least squares problem:
\begin{align}\label{eq:regulinHilbert} \hat g(t) = \argmin_{g\in\mathcal H_k} \sum_{t=1}^N (y(t) - (g*u)(t))^2 +
\gamma \|g\|_{\mathcal H_k}^2,
\end{align}
where $\|g\|_{\mathcal H_k}$ is the norm of $g$ in $\mathcal H_k$, and $\gamma>0$ is a regularization parameter and
controls the tradeoff between the data fit $\sum_{t=1}^N (y(t) -
(g*u)(t))^2$ and the regularization term $\|g\|_{\mathcal H_k}^2$.

We will discuss the issue of kernels in the next subsections. For the time being, we assume that a suitable kernel $k(t,s;\theta)$ has been designed, but its hyper-parameter $\theta$ is left to be determined. The current most widely used method to determine $\theta$ is the so-called empirical Bayes method \cite{PC15}. It first embeds the regularization in a Bayesian framework, and then estimates $\theta$ and possibly also the noise variance $\sigma^2$ by maximizing the marginal likelihood $p(Y|\eta)$, where $Y\in\R^N$ with $y(t)$ being its $t$th element, and $\eta$ could be $\theta$ or the concatenation of $\theta$ and $\sigma^2$. Specifically, we define $A(\theta)\in\R^{N\times N}$ with its $(t,s)$th element $A_{t,s}(\theta)$ defined by
\begin{align*}
a(t,s;\theta)=(k(t,\cdot;\theta)*u)(s),
A_{t,s}(\theta)=(a(\cdot,s;\theta)*u)(t)
\end{align*} and moreover, we let $\Sigma(\eta)= A(\theta)+\sigma^2I_N$, where $I_N$ is the $N$-dimensional identity matrix. Then we get \begin{align*}
 \minimize_{\eta\in\Gamma} Y^T\Sigma(\eta)^{-1}Y + \log\det \Sigma(\eta), \end{align*} where $\Gamma$ is a constraint set where we search for $\eta$.  When an estimate of $\eta$ is obtained, the solution
to (\ref{eq:regulinHilbert}) is
given by the representer theorem \cite[Theorem 3]{PDCDL14}:
\begin{align} \label{eq:solution2regulinHilbert}
\hat g(t)=\sum_{s=1}^N \hat c_{s} a(t,s;\theta),
\end{align} where $\hat c_{s}$ is the $s$th element of $\hat c=(A(\theta) + \sigma^2 I_N)^{-1}Y$.

\subsection{Existing Single Kernels}
So far the two mostly widely used single kernels are the stable spline
(SS) kernel \cite{PN10a} and the diagonal/correlated (DC) kernel
\cite{COL12a}. The SS kernel is defined as \begin{align}
&k^{\text{SS}}(t,s;\theta)= c\frac{1}{2}\exp(-\beta (t+s)-\beta\max\{t,s\})\nonumber\\&-c\frac{1}{6}\exp(-3\beta\max\{t,s\}),\ \theta=[c\ \ \beta]^T,\ c,\beta\geq0,\label{eq:SS_original}
\end{align}
and the DC kernel is defined as
 \begin{align}
\label{eq:DC} &k^{\text{DC}}(t,s;\theta)=c\lambda^{(t+s)/2}\rho^{|t-s|},\ \theta=[c\ \ \lambda]^T\\
\label{eq:TC} &k^{\text{TC}}(t,s;\theta)=c\min(\lambda^t,\lambda^{s}),\ \theta=[c\ \ \lambda\ \ \rho]^T
\\\nonumber&\quad \ c\geq0,0\leq\lambda<1, \left\{\begin{array}{cc}
                                             |\rho|\leq 1, & \text{ DT} \\
                                             0\leq\rho\leq 1, & \text{ CT}
                                           \end{array}\right.
  \end{align}
where the TC kernel is a special case of the DC kernel with $\rho=\sqrt{\lambda}$ and is also called the first order stable spline kernel.


\begin{Remark}\label{rmk:ct_DC_nonegativecoor}
It is worth to note that for $\rho<0$ and $|t-s|\not\in\mathbb N$, $\rho^{|t-s|}$ is complex and thus $k^{\emph{DC}}(t,s;\theta)$ with $\rho<0$ is not well defined for the CT case.
\end{Remark}

\subsection{Optimal Kernel and Stable Kernel}\label{sec:opt_kernel}

For the KRM, the optimal kernel in the sense of minimizing the mean square error (MSE) exists \cite{COL12a,PDCDL14} and motivates a general guideline to design a kernel. To state this result, we assume that the data has been generated by (\ref{eq:sys}) for a \emph{true} impulse response $g^0(t)$ and we let $\bar g^0$ and $\bar{\hat g}$ to represent any finite dimensional vector obtained by sampling $g^0(t)$ and its estimate $\hat g(t)$ at the same arbitrary time instants. Then the following result holds.
\begin{Lemma}[Optimal kernel, \cite{COL12a,PDCDL14}] Letting $\gamma=\sigma^2$. Then for the KRM, the MSE matrix \begin{align}
\text{MSE}(k) \triangleq \mathbb E\left[ (\bar{\hat g} - \bar g^0)(\bar{\hat g} - \bar g^0)^T\right],
\end{align} where $\mathbb E$ is the mathematical expectation, is minimized by the kernel \begin{align}\label{eq:OptKernel}
  k^{\emph{opt}}(t,s) = g^0(t)g^0(s),\quad t,s\in X
\end{align} in the sense that $\text{MSE}(k)-\text{MSE}(k^{\emph{opt}})$ is positive semidefinite for any positive semidefinite kernel $k$.
\end{Lemma}

The optimal kernel $k^{\text{opt}}$ cannot be applied in practice as it depends on the true impulse response $g^0(t)$. However, it motivates a general \emph{guideline} to design a kernel: \emph{let the kernel mimic the behavior of the optimal kernel, and moreover, the prior knowledge of the true impulse response should be used in the design of the kernel.}

For instance, if the LTI system is known to be stable, i.e., $g\in
\ell^1$ or $\mathcal L^1$, then the designed kernel $k$ should
reflect this, and a necessary condition to satisfy is that $k(t,t)$
should tend to 0 as $t$ goes to infinity. This observation basically
rules out the possibility to model the impulse response of stable
LTI systems with stationary kernels\footnote{Recall that a kernel
$k(t,s)$ with $t,s\in X$ is said to be stationary if $k(t,s)$ is a
function of $t-s$ for any $t,s\in X$.} and a rigorous proof for this
has been given in \cite[Lemma 8]{Dinuzzo12}. More generally, the
designed kernel should guarantee that its associated RKHS $\mathcal
H_k$ is a subspace of $\ell^1$ or $\mathcal L^1$ and a kernel that
has such property is said to be stable \cite{Dinuzzo12,PDCDL14}.
Sufficient and necessary conditions for a kernel to be stable exist
and are given below.

\begin{Lemma}[\cite{CDT06,Dinuzzo12}]\label{thm:stability} A positive semidefinite kernel $k$ is stable if and only if \begin{align}\label{eq:stabilitytest_sufnec}
\begin{array}{cc}
  \text{DT: } & \sum_{s=1}^\infty \left|\sum_{t=1}^\infty k(t,s)l(t)\right|
<\infty,\ \forall\ l(t)\in \ell^\infty
 \\
  \text{CT: } &   \int_0^\infty \left|\int_0^\infty k(t,s)l(t)dt\right|ds
<\infty,\ \forall\ l(t)\in \mathcal L^\infty
\end{array}
\end{align} where $\mathcal L^\infty$ and $\ell^\infty$ denote the space of bounded functions on $\R_{\geq0}$ and bounded sequences, respectively.
\end{Lemma}

\begin{Corollary}[\cite{Dinuzzo12,PDCDL14}]\label{coro:stability}  A positive semidefinite kernel $k$ is stable if
\begin{align}
\begin{array}{cc}
  \text{DT: } & \sum_{s=1}^\infty \left|\sum_{t=1}^\infty k(t,s)\right|
<\infty,
 \\
  \text{CT: } &   \int_0^\infty \left|\int_0^\infty k(t,s)dt\right|ds
<\infty.
\end{array}
\end{align}
\end{Corollary}

\subsection{Kernel Design}

Hereafter, we focus on the problem of kernel design. The problem is
stated as follows: for the given prior knowledge of the impulse
response to be identified, our goal is to design a kernel such that
the prior knowledge is embedded in the kernel. The answer should
depend on the type of the prior knowledge. We will consider two
types of prior knowledge and discuss how to design kernels
accordingly from two perspectives: a machine learning perspective
and a system theory perspective.

It should be noted that the LTI system (\ref{eq:sys}) is assumed to be stable, meaning that $g\in \ell^1$ or $\mathcal L^1$ becomes our first prior knowledge. In this regard, Theorem \ref{thm:stability} becomes a rule that the designed kernel should obey.

\begin{Remark}
Other than stability, the cases where the impulse response or the corresponding LTI system is known to have relative degree, to be monotonic,  smooth and to have delay, have been discussed in \cite{Dinuzzo12} and in particular, sufficient and/or necessary conditions for a kernel to have such properties are given. For instance, if a CT kernel $k$ is $m$-times continuously differentiable, then every function $h\in\mathcal H_k$ is $m$-times continuously differentiable \cite[Corollary 4.36, p. 131]{SC08}, \cite[Lemma 5]{Dinuzzo12}.
\end{Remark}

\section{A Machine Learning Perspective} \label{sec:ml_perspective}

We first show that both the SS kernel and the DC kernel belong to the class of the so-called amplitude modulated locally stationary (AMLS) kernel. Accordingly, we propose to treat the impulse response as a function, whose amplitude varies with a certain rate and decays exponentially to zero in the end, and moreover, design AMLS kernels to model the impulse response.

\subsection{Amplitude Modulated Locally Stationary Kernels}\label{sec:ecls}

Recall that a kernel $k(t,s)$ is said to be a locally stationary (LS)
kernel in the sense of Silverman \cite{Silverman57} if
\begin{align}\label{eq:ls_kernel}
  k(t,s) = k^{d}\left(\frac{t+s}{2}\right) k^{c}(t-s), \quad t,s\in X
\end{align} where $k^{d}\geq0$ and $k^{c}$ is a stationary kernel\footnote{Clearly, if $k^{d}$ is a positive constant, then the LS kernel reduces to a stationary kernel. }.
Motivated by the LS kernel, we introduce a kernel suitable for modeling impulse response, which is called the amplitude modulated locally stationary (AMLS) kernel.
\begin{Definition} A kernel $k(t,s)$ is said to be an amplitude modulated locally stationary kernel if
\begin{align}\label{eq:AMLS}
  k(t,s) &= k^{d}(t,s) k^{c}(t-s), \ k^{d}(t,s)=b(t)b(s),\\
  &\qquad\nonumber k^{c}(0)=1, \quad t,s\in X,
\end{align}
where $b(t)>0$ is bounded and $k^{c}$ is a stationary kernel.
\end{Definition}

It can be proved that the AMLS kernels have the following properties (proof given in the Appendix).

\begin{Proposition}\label{prop:AMLS} Consider the AMLS kernel (\ref{eq:AMLS}). Then the following results hold:
\begin{enumerate}

\item[(a)]  $k^{d}(t,s)$ is a rank-1 kernel\footnote{A kernel $k(t,s)$ with $t,s\in X$ is said to be a rank-1 kernel if for any $t_i,s_i\in X$, $i=1,\dots,n$ and for any $n\in\mathbb N$, the kernel matrix $K$, defined by $K_{i,j}=k(t_i,t_j)$, is a rank-1 matrix.} and moreover satisfy \begin{align}\label{eq:rank-1kernel_identity}
  k^{d}(t,s)=\sqrt{k^{d}(t,t)k^{d}(s,s)}.
\end{align}

\item[(b)] Let $g(t)$ be a stochastic process with zero mean and the AMLS kernel as the covariance function. Then $k^{c}(t-s)$ is the correlation coefficient between $g(t)$
and $g(s)$.

\item[(c)] Let $h(t)$ be a stochastic process with zero mean and $k^{c}(t-s)$ as the covariance function. Then the stochastic process $g(t)\triangleq b(t)h(t)$ has zero mean and the AMLS kernel as its covariance function.

\end{enumerate}
\end{Proposition}

\begin{Remark}\label{rmk:k2_bounded}
Since $k^{c}(t-s)$ is the correlation coefficient and $k^{c}(0)=1$, then $k^{c}(t-s)$ is bounded and moreover, $|k^{c}(t-s)|\leq 1$ for any $t,s\in X$.
\end{Remark}

For a function estimation problem, if a zero mean Gaussian processes with the AMLS kernel as the covariance function is chosen to model the function, then Proposition \ref{prop:AMLS} has the following implications:
\begin{itemize}

\item The stationary kernel $k^{c}(t-s)$ in (\ref{eq:AMLS}) accounts for the varying rate of the function.

Note that part (b) shows that $k^{c}(t-s)$ is the correlation
coefficient for $g(t)$, which implies that \begin{align*} &\mathbb E
(g(t)-g(s))^2\\ &\qquad= (b(t))^2+(b(s))^2 - 2 b(t)b(s)k^c(t-s).
\end{align*} The above equation shows that for any fixed $t,s\in X$, as $k^{c}(t-s)$ varies from 1
to $-1$, $(g(t)-g(s))^2$ tends to become larger, that is, $g(s)$
tends to vary more quickly away from $g(t)$. This observation also
holds for $h(t)$ as $k^{c}(t-s)$ is also the correlation coefficient
for $h(t)$.

\item The rank-1 kernel $k^{d}(t,s)$ in (\ref{eq:AMLS}) account for the change of the amplitude of the function.

Note that part (c) shows that the amplitude of $g(t)$ is modulated by the factor $b(t)$. If $h(t)$ does not converge to 0 as $t$ goes to $\infty$, then a suitable $b(t)$ can be chosen such that $g(t)$ does.

\end{itemize}

Interestingly, both the SS and DC kernels can be put in the form of
(\ref{eq:AMLS}).
Setting $\lambda=\exp(-\beta/2)$ and using the equality $\max\{t,s\}=(t+s+|t-s|)/2$ yields that the SS kernel (\ref{eq:SS_original}) is rewritten as
\begin{align}
\label{eq:SS}  &k^{\text{SS}}(t,s;c,\lambda)= c\lambda^{3(t+s)}\left(\frac{1}{2}\lambda^{|t-s|}-\frac{1}{6}\lambda^{3|t-s|}\right).
\end{align}
Then we identify,  for the SS kernel (\ref{eq:SS}),
\begin{equation}\label{eq:SS_decomp}\begin{aligned} k^{d}(t,s)=\frac{c}{3}\lambda^{3
(t+s)}, k^{c}(t-s) = \frac{3}{2}\lambda^{|t-s|}-\frac{1}{2}\lambda^{3|t-s|}, \end{aligned}\end{equation}
and for the DC kernel (\ref{eq:DC}), \begin{align}\label{eq:DC_decomp}
k^{d}(t,s) = c\lambda^{\frac{1}{2}(t+s)}, k^{c}(t-s) = \rho^{|t-s|}.
\end{align}
Moreover, one can prove the following result (proof given in the Appendix).
\begin{Proposition}\label{prop:ss&dc_amls} The SS kernel (\ref{eq:SS}) and the
DC kernel (\ref{eq:DC}) are AMLS kernels.
%
%
%
%
%
%
\end{Proposition}

\begin{figure}[!h]
\begin{center}
\includegraphics[width=3.4in, height=3in]{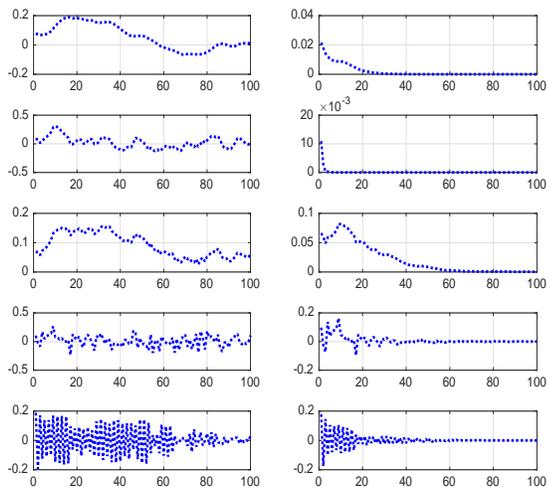}
\end{center}
\caption{Realizations of the DT zero mean Gaussian processes with
the SS kernel (\ref{eq:SS}) and the DC kernel (\ref{eq:DC}) as
covariance function. For each row, the left panel shows the
realizations of the DT zero mean Gaussian process with the IS kernel
$k^{c}(t-s)$ as covariance function and the right panel shows the
realization of the DT zero mean Gaussian process with the SS or DC
kernel as covariance function. The corresponding realizations for
the SS kernel (\ref{eq:SS}) are shown in the top two rows, which
correspond to $c=1$ and
$\lambda=0.9^{\frac{1}{2}},0.5^{\frac{1}{2}}$, respectively. The
corresponding realizations for the DC kernel (\ref{eq:DC}) are shown
in the bottom three rows, which correspond to $c=1$, $\lambda=0.9$
and $\rho=0.99,0,-0.99$, respectively.} \label{fig:exm1}
\end{figure}


Now we demonstrate for the SS and DC kernels the role of the rank-1 kernel $k^{d}$ and the stationary kernel $k^{c}$.

\begin{Example}\label{exm1}
As seen from the left panel of Fig. \ref{fig:exm1}, as $\lambda$
changes from $0.9^{\frac{1}{2}}$ to $0.5^{\frac{1}{2}}$, the
realizations of the DT zero mean Gaussian process with
$k^{c}(t-s;[c,\lambda]^T)$ in (\ref{eq:SS_decomp}) as the covariance
function varies more quickly, because $k^{c}(1;[1,
0.9^{\frac{1}{2}}]^T)> k^{c}(1;[1, 0.5^{\frac{1}{2}}]^T)$.
Similarly, as $\rho$ changes from $0.99$, to $0$ and to $-0.99$, the
realizations of the DT zero mean Gaussian process with
$k(t-s;[1,0.9,\rho]^T)$ in (\ref{eq:DC_decomp}) varies more quickly,
and especially for the case with $\rho=-0.99$, the realization tends
to change its sign at the next time instant.  The above observations
carry over to the right panel of Fig. \ref{fig:exm1}. Finally, the
realizations on the left panel of Fig. \ref{fig:exm1} do not go to
zero for large $t$ but the realizations on the right panel do,
because of $k^{d}(t,s)$.
\end{Example}

\begin{Remark}
Recall from Remark \ref{rmk:ct_DC_nonegativecoor} that the DC kernel (\ref{eq:DC}) with $\rho<0$ is only defined for the DT case but not for the CT case. Now we see that the DC kernel is not good to model quickly varying impulse responses for the CT case, which is also true for the SS kernel (\ref{eq:SS}).
\end{Remark}


\subsection{Construct AMLS Kernels for Regularized Impulse Response Estimation}

Proposition \ref{prop:ss&dc_amls} and its implications on the role
of the rank-1 kernel $k^{d}$ and the stationary kernel $k^{c}$
suggest a machine learning method to design kernels for regularized
impulse response estimation. If the impulse response is treated as a
function and the prior knowledge is about its decay and varying
rate, then AMLS kernels can be designed by choosing suitable rank-1
kernel $k^{d}$ and stationary kernel $k^{c}$ to account for the
decay and varying rate of the impulse response, respectively.

\subsubsection{Stationary Kernel $k^{c}(t-s)$ }\label{sec:ISkernel}

An important class of stationary kernels is the isotropic stationary (IS) kernel. Recall that a stationary kernel $k^{c}(t-s)$ is said to be an IS kernel if it depends on $|t-s|$. IS kernels have
been studied extensively in the literature in statistics and machine
learning, see e.g. \cite[Section 4.2.1]{RasmussenW:06}. There are
many choices of IS kernels that could be used instead of the ones in
(\ref{eq:SS_decomp}) and (\ref{eq:DC_decomp}).

Most of the IS kernels introduced in \cite[pages 83
-88]{RasmussenW:06} decay monotonically w.r.t. $|t-s|$ and are always positive. For
example, the squared exponential
(SE) kernel and the Mat\`{e}rn class of kernels:
\begin{align*}
               k^{c}(r) &= e^{-\beta r^2}, \ \beta>0,\ \  \text{``SE''}, \\
                k^{c}(r) &= \frac{2^{1-\nu}}{\Gamma(\nu)}\left(\beta \sqrt{2\nu}r
                \right)^\nu K^\nu \left(\beta \sqrt{2\nu}r
                \right),\\& \qquad\qquad \beta>0,\nu>0   \text{``Mat\`{e}rn''},
\end{align*} where $r=|t-s|$, $\Gamma(\cdot)$ is the Gamma function and $K^\nu$ is a modified Bessel function with order
$\nu$.

There are IS kernels that do \emph{not} decay monotonically, can take negative values, and can have the form of damped oscillation w.r.t. $|t-s|$.
For example, the kernel in \cite[p. 89]{RasmussenW:06}, \cite{Yaglom87}:
\begin{subequations}\begin{align}\label{eq:ctkernel_dampedoscillation}
k^{c}(r) = c(\alpha r)^{-\nu} J_\nu(\alpha r),\quad \alpha>0,\nu\geq
-1/2,
\end{align} where $r=|t-s|$, $c$ is a scalar such that $k^{c}(0)=1$, and $J_\nu(\cdot)$ is the Bessel function of the first kind  with order
$\nu$. The kernel (\ref{eq:ctkernel_dampedoscillation}) is defined for any $r\geq0$ and can thus be used for both CT and DT impulse response estimation. In particular, for $\nu=-1/2$ and $\nu=1/2$, the kernel (\ref{eq:ctkernel_dampedoscillation}) takes the following form
\begin{align}\label{eq:ctkernel_dampedoscillation_special}
k^{c}(r) &= \cos(\alpha r),\ \alpha>0,\\
k^{c}(r) &= \frac{\sin(\alpha r)}{\alpha r},\ \alpha>0.
\end{align}
\end{subequations}

\subsubsection{Rank-1 Kernel $k^{d}(t,s)$}

The design of $k^{d}(t,s)$ is equivalent to that of the strictly
positive function $b(t)$ in (\ref{eq:AMLS}). The bottom line is that
$b(t)$ should ensure that the designed AMLS kernel (\ref{eq:AMLS})
is stable, i.e, $\mathcal H_k$ is a subspace of $\ell^1$ or
$\mathcal L^1$. Our following result gives a characterization of the
stability of the AMLS kernel (proof given in the Appendix).

\begin{Proposition}\label{thm:amls_stability} Consider the AMLS kernel (\ref{eq:AMLS}). Then the following results hold.
\begin{itemize}
  \item[(a)] If
$b(t)\in\ell^1$ for the DT case, and $b(t)\in\mathcal L^1$ for the CT case, then the AMLS kernel (\ref{eq:AMLS}) is stable.

\item[(b)] Assume that there exists a sequence of positive numbers $\lambda_i$ and linearly independent functions $\phi_i(t)$ defined on $X$ such that \begin{align}\nonumber
k^{c}(t-s)& = \sum_{i=1}^\infty \lambda_i \phi_i(t)\phi_i(s),\\&\quad t,s\in X, \lambda_i >0, i=1,\cdots,\infty, \label{eq:k2_mercer}
\end{align}  where the convergence is absolute and uniform on $Y_1\times Y_2$ with $Y_1,Y_2$ being any compact subsets of $X$.
If the AMLS kernel (\ref{eq:AMLS}) is stable, then there exists no $\epsilon >0$ such that $b(t)\geq \epsilon$ for all $t\in X$.
\end{itemize}
\end{Proposition}

Since the $b(t)$ in (\ref{eq:SS_decomp}) and (\ref{eq:DC_decomp})
are exponential decay functions and clearly satisfy $b(t)\in\ell^1$
or $\mathcal L^1$, the SS and DC kernels are stable by Proposition
\ref{thm:amls_stability}.

\begin{Remark}\label{rmk:k2_mercer} It is reasonable to check wether the condition (\ref{eq:k2_mercer}) is too strong to be satisfied. In fact, the series expansion (\ref{eq:k2_mercer})
can be obtained by Mercer's Theorem \cite{Mercer09,Hochstadt73,Sun05}.
For instance, a sufficient condition for (\ref{eq:k2_mercer}) to hold is given in \cite{Sun05}. To state this result, we define the kernel section of $k^{c}$ at a fixed $s\in X$ as $k^{c}_s\triangleq k^{c}(t-s)$ and
we let $\mathcal L_2(X,\mu)$ denote the space of functions $f:X\rightarrow\R$ such that $\int_X |f(t)|^2d\mu(t)<\infty$, where $\mu$ is a nondegenerate Boreal measure on $X$. Then (\ref{eq:k2_mercer}) holds if  $k^{c}(t-s)$ is continuous, $k^{c}_s \in \mathcal L_2(X,\mu)$ and
\begin{align}
\int_X\int_X (k^{c}(t-s))^2 d\mu(t)d\mu(s)<\infty.
\end{align} It is easy to check that many stationary kernels satisfy the above sufficient condition, e.g., the SE kernel and the $k^{c}$ in (\ref{eq:SS_decomp}) and (\ref{eq:DC_decomp}). 
\end{Remark}

\begin{Remark}\label{rmk:stability}
It follows from (\ref{eq:proof_int1}) that if there exists an $\epsilon>0$ such that $|h(t)|\geq\epsilon$ for all $t\geq0$, then $b(t)\in\mathcal L^1$, implying that $b(t)\in\mathcal L^1$ is also necessary for the stability of the AMLS kernel (\ref{eq:AMLS}). The above observations show that under the assumption that the AMLS kernel (\ref{eq:AMLS}) is stable, the result we can draw on the properties of $b(t)$ is determined by the property of $\mathcal H_{k^{c}}$.
\end{Remark}

\begin{Remark}
The reason why we force $b(t)>0$ for $t\in X$ is because for a
kernel in the form of $k(t,s)=k^{d}(t,s)k^{c}(t-s)$, we expect the
two kernels $k^{d}$ and $k^{c}$ have somewhat independent role. As
shown above, this idea eases the kernel design and the corresponding
analysis. If this idea is not taken, then we can design more general
$b(t)$ and even more general $k^{d}$. For instance, we could allow
$b(t)$ to be arbitrary real-valued function and we could also allow
$k^{d}$ to be the more general exponentially convex (EC)
kernel\footnote{For the LS kernel (\ref{eq:ls_kernel}), if $k^{d}$
is also a kernel, then $k^{d}$ is called an exponentially convex
(EC) kernel \cite{Loeve78} and in this case $k$ is called an
exponentially convex locally stationary (ECLS) kernel
\cite{Silverman57}.} and design accordingly the ECLS kernel
\cite{CL15a} instead of the AMLS kernel. The price to pay is that
the role of $k^{d}$ and $k^{c}$ would become obscure and in
particular, $k^{d}$ would not describe the decay rate and $k^{c}$
would not be the correlation coefficient of the underlying Gaussian
process.
\end{Remark}

\section{A System Theory Perspective}\label{sec:st_perspective}

Instead of simply treating the impulse response as a function, we
now associate the impulse response with an LTI system which is
stable and may be overdamped, underdamped, have multiple distinct
time constants and resonant frequencies, and etc., and our goal is
to design kernels to embed this kind of prior knowledge.

\subsection{Sketch of The idea}


Suppose that the prior knowledge of an LTI system is embedded in a
stochastic process $g(t)$ that is used to model the corresponding
impulse response. Then following the guideline to design kernels in
Section \ref{sec:opt_kernel} that is to mimic the behavior of the
optimal kernel (\ref{eq:OptKernel}), we should design the kernel as
\begin{align}\label{eq:si_kernel_idea} k(t,s)=\mathbb{C}ov (
g(t),g(s)),\end{align} where $\mathbb{C}ov ( g(t), g(s))$ is the
covariance between $g(t)$ and $ g(s)$. Now the problem of kernel
design becomes ``how to embed the prior knowledge of an LTI system
in a stochastic process $g(t)$ that is used to model the
corresponding impulse response''?

\begin{figure}[!t]
\begin{center}
\includegraphics[width=3in, height=0.9in]{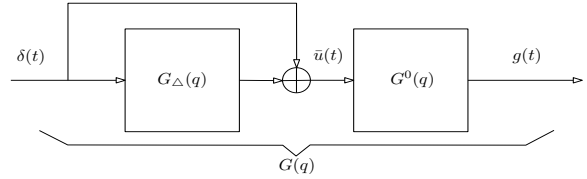}
\end{center}
\caption{The block diagram for the multiplicative uncertainty paradigm in robust control. The single-input-single-output system $G(q)=G^0(q)(1+G_\triangle(q))$ consists of two parts: the nominal part $G^0(q)$ and the uncertainty part $G_\triangle(q)$, where $q$ is the forward shift operator and the differential operator for the DT and CT case, respectively. The real-valued signals $\delta(t),\bar u(t)$ and $g(t)$ are the impulsive input, the input of $G^0(q)$ and the output of $G(q)$, respectively. }
\label{fig:multiuncertainty}
\end{figure}

A natural way to tackle the above question is by using simulation.
To this goal, it is useful to employ the multiplicative uncertainty
configuration in robust control, see e.g., \cite{ZhouDG:96}, as
shown in Fig. \ref{fig:multiuncertainty}. Here, the nominal model
$G^0(q)$ is used to embed the prior knowledge on the LTI system to
be identified, the uncertainty $G_\triangle(q)$ is assumed to be
stable and finally, the multiplicative uncertainty configuration is
used to take into account both the nominal model and the
uncertainty, and is simulated with an impulsive input to get a
zero-mean Gaussian process to model the impulse response $g(t)$.
This idea leads to the system theory method to design kernels and we
call such kernels the simulation induced (SI) kernels.

\begin{Remark}\label{rmk:multiplicatieuncertain}
It is worth to note that the problem of marrying system
identification and robust control is not new and has been studied
over two decades in the system identification community, see e.g.,
\cite{NG95,Ljung:99f}.  Both the additive uncertainty configuration
and the multiplicative uncertainty configuration have been used and
many results have been obtained, see e.g.,
\cite{GoodwinGN:92,NG95,Ljung:99f,Goodwin:02,RGL02,LGAC15}. In
particular, \cite{GoodwinGN:92,NG95,Goodwin:02,RGL02,LGAC15} used
the nominal model $G_0(q)$ and the uncertainty $G_\triangle(q)$ to
represent a model estimate and the corresponding model error,
respectively. Here the nominal model $G_0(q)$ and the uncertainty
$G_\triangle(q)$ are used to embed the prior knowledge of the LTI
system and describe the corresponding uncertainty, respectively.
\end{Remark}

\subsection{Simulation Induced Kernel}

More specifically, the prior knowledge is embedded in $G^0(q)$ or equivalently, in its state-space model \begin{subequations}\label{eq:state-space_model}\begin{align}
& &z(t+1) = Az(t) + B\bar u(t)\\
&G^0(q): &\dot z(t)= Az(t) + B\bar u(t)\\
& &\qquad z(0)\sim \N(0,Q)\\
&  &g(t) = Cz(t)+D\bar u(t)
\end{align}\end{subequations} where $A,B,C,D$, and $Q$ have compatible dimensions, $\bar u(t)$ and $g(t)$ are the input and output of $G^0(q)$, respectively. In what follows, we will use the quintuple $(A,B,C,D,Q)$ to represent the state-space model (\ref{eq:state-space_model}).

Since the uncertainty $G_\triangle(q)$ is stable, the simplest way
to model the impulse response of $G_\triangle(q)$ with a stochastic
process $l(t)$, is to have
\begin{align}\label{eq:uncertainty_impulse} G_\triangle(q): \quad
l(t) = b(t) w(t),\quad t\in X,
\end{align}  where $w(t)$ is a white Gaussian noise with zero mean and unit variance (independent of $z(0)$), and $b(t)>0$ and $b(t)\in\ell^1$ or $\mathcal L^1$ for the DT and CT case, respectively.
Clearly, $l(t)$ is a zero mean Gaussian process with the AMLS kernel
$b(t)b(s)\delta(t-s)$ as the covariance function, where
$\delta(\cdot)$ is the Dirac Delta function and the AMLS kernel
$b(t)b(s)\delta(t-s)$ is stable by Proposition
\ref{thm:amls_stability}.

\begin{figure}[!t]
\begin{center}
\includegraphics[width=3in, height=0.9in]{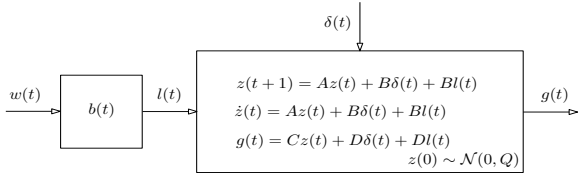}
\end{center}
\caption{The block diagram for the SI kernel (\ref{eq:si_kernel}). }
\label{fig:si_kernel}
\end{figure}

Finally, we simulate the system $G(q)$ in Fig. \ref{fig:multiuncertainty} with the impulsive input $\delta (t)$ and noting (\ref{eq:state-space_model}), (\ref{eq:uncertainty_impulse}) and $\bar u(t)=\delta(t)+l(t)$, we obtain the simulation induced (SI) kernel:
\begin{subequations}\label{eq:si_kernel}
\begin{align}
&z(t+1) = Az(t) + B\delta(t) + Bb(t)w(t),\\
&\dot z(t)= Az(t) + B\delta(t) + Bb(t)w(t),\label{eq:si_kernel_ct_sde}\\
&g(t) = Cz(t)+ D\delta(t) + Db(t)w(t),\\
&\qquad \qquad z(0)\sim \mathcal N(0,Q)\\
&k^{\text{SI}}(t,s)=\mathbb Cov(g(t),g(s)),\end{align}
\end{subequations}
whose block diagram is shown in Fig. \ref{fig:si_kernel}.  From linear system theory, e.g., \cite{Chen99}, the formal expression of the SI kernel (\ref{eq:si_kernel}) is available. For the DT case, it is
\begin{subequations}\label{eq:si_kernel_formalexpression}
\begin{align}\label{eq:si_kernel_dt} &k^{\text{SI}}(t,s)
=CA^tQ(A^{s})^TC^T + D^2
b(t)b(s)\delta(t-s)\nonumber
\\\nonumber& +Db(t)\sum_{k=0}^{s-1}\delta(t-k)b(k)B^T(A^{s-1-k})^TC^T
\\\nonumber& +Db(s)\sum_{k=0}^{t-1}\delta(s-k)b(k)B^T(A^{t-1-k})^TC^T
\\&+
\sum_{k=0}^{\min\{t,s\}-1} b(k)^2
CA^{t-1-k}BB^T(A^{s-1-k})^TC^T\end{align}
For the CT case with $D=0$, it is
\begin{align}\label{eq:si_kernel_ct} &k^{\text{SI}}(t,s)=  C
e^{At}Q(e^{As})^TC^T +  \nonumber\\ & C\int_0^{\min\{t,s\}} b^2(\tau)
e^{A(t-\tau)} B B^T (e^{A(s-\tau)})^T
d\tau C^T
\end{align}
\end{subequations}

\begin{Remark}
It should be noted that both DT and CT cases are considered in (\ref{eq:state-space_model}) and (\ref{eq:si_kernel}). For the CT case, (\ref{eq:si_kernel_ct_sde}) could be more rigorously written as an It\^{o} stochastic differential equation (SDE) as in \cite[eq. (9)]{CL14}. However, we decide to take the same point of view as in \cite{GL00} to use (\ref{eq:si_kernel_ct_sde}) instead in order to save the space for the introduction of SDE stuff, which is only used to derive (\ref{eq:si_kernel_ct}).
\end{Remark}

\begin{Remark}
The SI kernels (\ref{eq:si_kernel_formalexpression}) may or may not have closed form expressions. The related computational difficulty and cost depend on whether or not the SI kernels (\ref{eq:si_kernel_formalexpression}) have closed form expressions. If they do, then the computation would be easier and similar to the SS and DC kernels. If they do not, the computation of the hyper-parameter estimate and the regularized impulse response estimate would become more demanding. In this regard, the particle filtering based technique for nonlinear state-space model identification in \cite{SchonWN:11} could be adopted. However, the technique in \cite{SchonWN:11} cannot be applied trivially as the measurement output (\ref{eq:sys}) for the state space model (\ref{eq:si_kernel}) does not depend on the current state $z(t)$ solely but also the past state due to the presence of convolution. More details will be reported in an independent paper.
\end{Remark}

Note that $g(t)$ in (\ref{eq:si_kernel}) is a Gaussian process with the impulse response of the LTI system $(A,B,C,D,Q)$ as its mean and the SI kernel $k^{\text{SI}}(t,s)$ as its covariance function. If $\delta(t)$ is set to 0 in (\ref{eq:si_kernel}), $g(t)$ in (\ref{eq:si_kernel}) is a zero mean Gaussian process with the SI kernel $k^{\text{SI}}(t,s)$ as its covariance function. In what follows, when considering kernel design, we will set $\delta(t)$ to zero for convenience.

Interestingly, the AMLS kernel (\ref{eq:AMLS}) is closely related with the SI kernel (\ref{eq:si_kernel}), and in fact many AMLS kernels such as the SS and DC kernels can be put in the form of
(\ref{eq:si_kernel}) and are thus SI kernels. To state this result, recall that the power spectral density of $k^{c}(t-s)$, denoted by $\Psi(\omega)$, is defined as
\begin{align} \Psi(\omega)
=\left\{ \begin{array}{cc}
           \sum_{\tau=-\infty}^{+\infty} k^{c}(\tau) e^{-i\omega \tau},
 & \text{DT}, \\
           \int_{-\infty}^{+\infty} k^{c}(\tau) e^{-i\omega \tau} d\tau,
 & \text{CT}.
         \end{array}
\right.
\end{align}

\begin{Proposition}\label{prop:amls_sikernel} Consider the AMLS kernel (\ref{eq:AMLS}).
Assume that $k^{d}(t,s)=c\lambda^{t+s}$ with $c\geq0$ and $0\leq\lambda<1$and  $\Psi(\omega)$ is a proper rational function of
$e^{i\omega}$ or $\cos(\omega)$ for the DT case, and $\omega$ for the CT case, respectively. Then the AMLS kernel (\ref{eq:AMLS}) can be put in the form of (\ref{eq:si_kernel}) and thus is SI kernel.

\end{Proposition}

\begin{Proposition}\label{prop:si_kernel4ssdc} Consider the SS kernel (\ref{eq:SS}) and the DC kernel (\ref{eq:DC}). Then the following results hold:
\begin{itemize}
\item For the DT case, let \begin{align*}&\bar a=\sqrt{1+\lambda^2-\sqrt{1+\lambda^2+\lambda^4}},\\
&\bar b=\sqrt{1+\lambda^2+\sqrt{1+\lambda^2+\lambda^4}}.\end{align*} Then the SS kernel is in the form of
(\ref{eq:si_kernel}) with
\begin{align}\label{eq:SS_ssrform_dt}
\nonumber&A=\lambda^3\left[
            \begin{array}{cc}
              \lambda & 0 \\
              0 & \lambda^3 \\
            \end{array}
          \right], B= \left[
         \begin{array}{c}
           \lambda^3 \\
           \lambda^3 \\
         \end{array}
       \right],Q= \frac{c}{3}
\left[
  \begin{array}{cc}
    \frac{1}{1-\lambda^2} & \frac{1}{1-\lambda^4} \\
    \frac{1}{1-\lambda^4} & \frac{1}{1-\lambda^6} \\
  \end{array}
\right]\\\nonumber &C=\sqrt{\frac{(1-\lambda^2)^3}{2}}\left[
                        \begin{array}{cc}
                          \frac{\bar a+\bar b\lambda}{1-\lambda^2} & \frac{-\lambda^2(\bar a+\bar b\lambda^3)}{1-\lambda^2}
                        \end{array}
                      \right]\\ & D= \bar b\sqrt{\frac{(1-\lambda^2)^3}{2}}, b(t)
                      =\left(\frac{c}{3}\right)^{\frac{1}{2}}\lambda^{3t}.
\end{align}
The DC kernel is in the form of (\ref{eq:si_kernel}) with
\begin{align} \label{eq:DC_ssrform_dt} \nonumber &A= \lambda^{\frac{1}{2}}\rho, B=\lambda^{\frac{1}{2}},C=\rho(1-\rho^2)^{\frac{1}{2}}, Q=\frac{c}{1-\rho^2},\\&D=(1-\rho^2)^{\frac{1}{2}},b(t)=c^{\frac{1}{2}}\lambda^{\frac{t}{2}}.\end{align}

\item For the CT case, let $\lambda=\exp(-\beta/2)$ and $\rho=\exp(-\alpha)$. Then the SS kernel is in the form of
(\ref{eq:si_kernel}) with
\begin{align}\nonumber&A=\left[
            \begin{array}{cc}
              -\frac{3}{2}\beta & 1 \\
              -\frac{3}{4}\beta^2 & -\frac{7}{2}\beta \\
            \end{array}
          \right], B= \left[
         \begin{array}{c}
           0 \\
           1 \\
         \end{array}
       \right],Q= \frac{c}{3}
\left[
  \begin{array}{cc}
    \frac{1}{3\beta^3} & 0\\
    0 & \frac{1}{4\beta} \\
  \end{array}
\right]\\\label{eq:SS_ssrform_ct}
 &C=\left[
                        3^{\frac{1}{2}}\begin{array}{cc}
                          \beta^{\frac{3}{2}} & 0
                        \end{array}
                      \right], D = 0, b(t)
                      =\left(\frac{c}{3}\right)^{\frac{1}{2}}e^{-\frac{3}{2}\beta t}.
\end{align}
The DC kernel is in the form of (\ref{eq:si_kernel}) with
\begin{align}\nonumber
 &A= -\alpha-\frac{1}{4}\beta, B=1,C=(2\alpha)^{\frac{1}{2}}, Q=\frac{c}{2\alpha},\\&D=0,b(t)=c^{\frac{1}{2}}e^{-\frac{\beta}{4}t}.
\label{eq:DC_ssrform_ct}\end{align}

\end{itemize}
\end{Proposition}

Proposition \ref{prop:si_kernel4ssdc} enhances our understanding about the SS and DC kernels and in particular, the prior knowledge embedded in the two kernels from a system theory perspective is as follows:
\begin{itemize}
\item[1)] For the SS kernel, the corresponding nominal model $G^0(q)$ in Fig. \ref{fig:multiuncertainty} is a second order
system. For the DC kernel, the corresponding nominal model $G^0(q)$
is a first order system. For both the SS and DC kernels and for the
CT case, $G^0(q)$ has negative real poles corresponding to
overdamped impulse responses. For the DT case, the two poles for the
SS kernel are real positive and correspond to impulse response
without oscillation, and the pole for the DC kernel can be positive
or negative (depending on $\rho$) and corresponds to impulse
response without or with oscillation. The pole for the TC kernel is
positive (equal to $\lambda$) and corresponds to impulse response
without oscillation.

\item[2)] For both the SS and DC kernels, the decay rate of the impulse response of the uncertainty $G_\triangle(q)$ in Fig. \ref{fig:multiuncertainty} are all described by exponential decay functions.
\end{itemize}

\begin{Remark}\label{rmk:statespace}
It is worth to note that to study the link between a kernel and its
state-space model realization has a long history, see e.g.,
\emph{\cite{Astrom:70,GL00}}. This link is important because it
opens the body of a kernel from a system theory perspective and
accordingly helps to understand the underlying behavior of a kernel.
\end{Remark}

\subsection{Stability of SI kernels}

For the SI kernel (\ref{eq:si_kernel_formalexpression}), we
provide the following sufficient condition to guarantee its stability.

\begin{Proposition}\label{prop:si_kernel_stability_suf}
Consider the SI kernel (\ref{eq:si_kernel_formalexpression}).
\begin{subequations}\label{eq:si_kernel_stablityassumption}
\begin{itemize}
\item For the DT case, assume
that $A$ has distinct eigenvalues and moreover, $A$ is stable, i.e., $A$ has
all eigenvalues strictly inside the unit circle.
Assume further that
\begin{align}\label{eq:si_kernel_dt_stablityassumption}
b(t)\leq \bar c |\bar\lambda|^{\frac{t}{2}}
\end{align} for some $\bar c>0$, where $\bar\lambda$
is the eigenvalue of $A$ with the largest modulus.
Then the SI kernel (\ref{eq:si_kernel_dt}) is
stable.

\item For the CT case, assume
that $A$ has distinct eigenvalues and moreover $A$ is stable, i.e., $A$ has
all eigenvalues with strictly negative real parts. Assume further that
\begin{align}\label{eq:si_kernel_ct_stablityassumption}
b(t)\leq \bar ce^{-\frac{1}{2}|\emph{Re}(\bar \lambda)| t}
\end{align} for some $\bar c>0$, where $\bar\lambda$
is the eigenvalue of $A$ with the largest real part and $\emph{Re}(\bar \lambda)$ is the real part of $\bar \lambda$.
Then the SI kernel (\ref{eq:si_kernel_ct}) is
stable.

\end{itemize}
\end{subequations}

 \end{Proposition}

\begin{Remark} The conditions (\ref{eq:si_kernel_stablityassumption}) are sufficient but not necessary. For instance, consider the DT case. The SS kernel (\ref{eq:SS_ssrform_dt}) satisfies (\ref{eq:si_kernel_dt_stablityassumption}) but the DC kernel (\ref{eq:DC_ssrform_dt}) does not. It is possible to derive a slower upper bound for $b(t)$. Another intension to have Proposition \ref{prop:si_kernel_stability_suf} is to give some guidelines about the relations between $b(t)$, $A$ and the stability of SI kernel: in order to guarantee stability of the SI kernel, the impulse response of the uncertainty $G_\triangle(q)$ (\ref{eq:uncertainty_impulse}), i.e., $b(t)$ should decay a bit faster than the slowest mode of the nominal model $G^0(q)$.

\end{Remark}

\subsection{Maximum Entropy Property of SI Kernels}

Since the prior knowledge is never complete, it is worth to note
Jaynes's maximum entropy (MaxEnt) rationale \cite{Jaynes82} to
derive from incomplete prior knowledge the optimal statistical prior
distribution. Jaynes's idea is to formulate a MaxEnt problem with
respect to the prior, and then solve the problem to obtain the
optimal prior. The constraints of the problem describes the prior
knowledge (in the MaxEnt sense) of the underlying stochastic process
(the system) to be identified. Interestingly, the SI kernel lends
itself easily to a MaxEnt interpretation, leading to a new facet to
understand the underlying behavior of the SI kernel.

%

Recall that the differential entropy $H(X)$ of a real-valued
continuous random variable $X$ is defined as $
H(X)=-\int_S{p(x)\log{p(x)} d x}$, where $p(x)$ is the
probability density function of $X$ and $S$ is the support set of
$X$. Then we prove the next MaxEnt interpretation of SI kernels\footnote{Let $X_1,X_2$ be two jointly distributed random variables. Then the joint differential entropy of $H([X_1,X_2]^T)$ is simply written as $H(X_1,X_2)$ below. }.

\begin{Proposition}\label{thm:maxent4si_kernel} Consider the SI kernel (\ref{eq:si_kernel}) with (\ref{eq:si_kernel_dt}).
\begin{itemize}

\item For the case $D\neq 0$, define
\begin{align}\label{eq:maxent4ssm_intmf_nonzero}
f(t)&=\frac{\bar g(t)-CA^t \bar z(0) - \sum_{k=0}^{t-1}CA^{t-1-k}B
b(k) f(k)}{Db(t)}\nonumber\\
&\qquad \bar g(t)\in\R,\bar z(0)\in\R^n, t=0,1,\cdots,s
\end{align}
Then for any $s\in\mathbb N$, the Gaussian process $g(t)$ in (\ref{eq:si_kernel}) is the
solution to the MaxEnt problem
\begin{subequations}\label{eq:maxent4ssm_Dnonzero}
\begin{align}
&\maximize_{\bar z(0),\bar g(t)} \quad H(\bar z(0),\bar g(0),\bar g(1),\cdots,\bar g(s))\\
&\mbox{subject to}\nonumber  \\
&\quad \mean{\bar z(0)}=0,\mean{\bar g(t)}=0, \nonumber\\&\quad
\cov{\bar z(0)} = Q, \var{f(t)}= 1, t =
0,\cdots,s\label{eq:maxent4ssm_Dnonzero_constr}
\end{align} where $\cov{\bar z(0)}$ is the covariance matrix of $z(0)$ and $\var{f(t)}$ is the variance of $f(t)$.
\end{subequations}

\item For the case $D=0$, assume that $CB\neq0$ and define \begin{align}
\label{eq:maxent4ssm_intmf_zero}
f(t-1)&=\frac{\bar g(t)-CA^t \bar z(0) - \sum_{k=0}^{t-2}CA^{t-1-k}B
b(k)f(k)}{CBb(t-1)}\nonumber\\
&\qquad\bar g(t)\in\R,\bar z(0)\in\R^n, t=0,\cdots,s-1
\end{align}
Then for any $s\in\mathbb N$, the Gaussian process $g(t)$ in (\ref{eq:si_kernel}) is the
solution to the MaxEnt problem
\begin{subequations}\label{eq:maxent4ssm_Dzero}
\begin{align}
&\maximize_{\bar z(0),\bar g(t)} \quad H(\bar z(0),\bar g(1),\cdots,\bar g(s))\\
&\mbox{subject to}\nonumber  \\
&\quad \mean{\bar z(0)}=0,\mean{\bar g(t)}=0, \nonumber\\&\quad
\cov{\bar z(0)} = Q, \var{f(t)}= 1, t = 0,\cdots,s-1
\label{eq:maxent4ssm_Dzero_constr}
\end{align}
\end{subequations}

\end{itemize}

\end{Proposition}


\begin{Corollary}
\label{coro:maxent4si_dc} For any $s\in\mathbb N$, the
zero mean Gaussian process $g(t)$ with the DC kernel (\ref{eq:DC}) defined on $\mathbb N\times \mathbb N$ as
its covariance is the solution to the MaxEnt problem
\begin{equation}\label{eq:maxent4dc}
\begin{aligned}
&\maximize_{\bar g(t)} \quad H(\bar g(0),\bar g(1),\cdots,\bar g(s))\\
&\mbox{subject to}\quad   \mean{\bar g(t)}=0, t=0,\cdots, s,
\var{\bar g(0)}=c,\\& \var{\bar g({t})-\lambda^{1/2}\rho \bar
g({t-1})}= c(1-\rho^2)\lambda^{t}, t=1,\cdots, s
\end{aligned}
\end{equation}
\end{Corollary}

\begin{Remark}
When $\rho=\lambda^{1/2}$, the DC kernel (\ref{eq:DC}) becomes
the TC kernel (\ref{eq:TC}). For the TC kernel defined on $\mathbb N\times \mathbb N$, (\ref{eq:maxent4dc}) becomes
\begin{equation}\label{eq:maxent4tc}
\begin{aligned}
&\maximize_{\bar g(t)} \quad H(\bar g(0),\bar g(1),\cdots,\bar g(s))\\
&\mbox{subject to}\quad   \mean{\bar g(t)}=0, t=0,\cdots, s,
\var{\bar g(0)}=c,\\& \var{\bar g({t})-\lambda \bar g({t-1})}=
c(1-\lambda)\lambda^{t}, t=1,\cdots, s
\end{aligned}
\end{equation} which is different from the MaxEnt interpretation
in \cite{CACCLP16}. It shall be noted that both interpretation are
correct but derived in different ways: the MaxEnt problems are
different but have the same optimal solution.

\end{Remark}

\begin{Remark}\label{rmk:DC-completion}
By using Corollary \ref{coro:maxent4si_dc} and the trick in
\cite[Theorem 2]{CACCLP16}, it is possible to derive a more concise
proof for \cite[Proposition IV.1]{CCL17} which shows that, the fact
that the DC kernel has tridiagonal inverse can be given a MaxEnt
covariance completion interpretation.
\end{Remark}

\begin{Remark}
The CT case is not discussed here because according to our best knowledge the entropy is not well defined for CT stochastic processes.
\end{Remark}

\subsection{Markov Property of SI Kernels}

As shown in \cite{Carli14,MSS16,CCL17,CACCLP16}, the kernel matrix
of DC kernel (\ref{eq:DC}) has tridiagonal inverse. Here, we further
show that the Gaussian process with the DC kernel as its covariance
function is also a Markov process with order 1 and moreover, we are
able to design SI kernels which correspond to more general Gaussian
Markov processes and have banded\footnote{A real symmetric matrix
$A$ with dimension $n> m+1$ is called an $m-$\emph{band} matrix if
$A_{i,j}=0$ for $|i-j|>m$.}inverses of their kernel matrices.


First, recall from e.g., \cite[Appendix B]{RasmussenW:06} that a Gaussian Markov
process is a stochastic process that is both a Gaussian process and
a Markov process. A well-known instance is the DT
autoregressive process of order $p$:
\begin{subequations}\label{eq:gmp}
\begin{align}\label{eq:gmp_1}
& x(t+1) = \sum_{k=0}^{p-1} a(t,k)x(t-k) + b(t+1)w(t+1)\\
&\text{or, }x(t+1) = \sum_{k=0}^{p-1} a(t,k)x(t-k) +
b(t)w(t)\label{eq:gmp_2}
\end{align}
\end{subequations}
where $t\in\mathbb N$, $x(t),a(t,k),b(t)\in\R$, $x(0)$ is Gaussian
distributed and assumed to be independent of $w(t)$, a zero mean
white Gaussian noise with unit variance.  The stochastic process (\ref{eq:gmp}) is a Markov process with order $p$ since $x(t+p+1)$ only depends on $x(t+p),\cdots,x(t+1)$ given the history $x(s)$ with $s\leq t+p$.

\begin{Proposition}\label{prop:dc_markov}
Consider a zero mean DT Gaussian process $g(t)$
with the DC kernel (\ref{eq:DC}) as its covariance. Then $g(t)$ with $t\in\mathbb N$ can be put in the form of
\begin{align}\label{eq:dc_gmp}
g(t+1) = \lambda^{\frac{1}{2}}\rho g(t) + (1-\rho^2)^{\frac{1}{2}}c^{\frac{1}{2}}\lambda^{\frac{t+1}{2}}w(t+1)
\end{align}
and thus a Markov process with
order 1, and moreover, its kernel matrix has a $1$-band matrix inverse.
\end{Proposition}

\begin{Remark}
Interestingly, (\ref{eq:dc_gmp}) can also be derived from  (\ref{eq:maxent4dc}), i.e., from the MaxEnt property of the DC kernel.
\end{Remark}

It is possible to construct more general SI kernels with Markov
property.

\begin{Proposition}\label{prop:markov4si_kernel}
Consider the DT SI kernel (\ref{eq:si_kernel}) with $(A,B,C,D)$ being a realization of $G^0(q)$ which is
an $n$th order DT system
\begin{align}\label{eq:highordcanonical}
G^0(q) = \frac{q^{n-1}\bar b}{q^n+a_1 q^{n-1}+\cdots+a_n},
\end{align} where $\bar b, a_1,\dots,a_n\in \R$.
Then the Gaussian process $g(t)$ in (\ref{eq:si_kernel}) with any $b(t)>0$ and positive semidefinite $Q$ is also a
Markov process with order $n$ and thus the SI kernel has an $n$-band
matrix inverse.
\end{Proposition}

\begin{figure}[!t]
\begin{center}
\includegraphics[width=2in, height=1.5in]{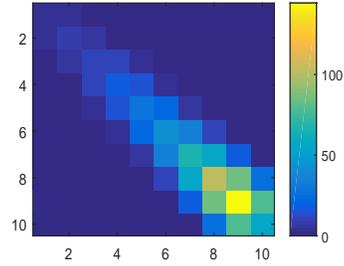}
\end{center}
\caption{Scaled image of $K^{-1}$ in Example \ref{exm:bandedinverse}. When generating $K$, we choose $\bar b=1$, $a_1=0.5$, $a_2=0.9$, $b(t)=0.8^t$, and $Q=I_2$. The image is drawn by using \texttt{imagesc} in MATLAB, where the colder the color the smaller the element of $K^{-1}$.}
\label{fig:bandedinverse}
\end{figure}

For illustration, we consider an example.
\begin{Example}\label{exm:bandedinverse}
Consider the DT SI kernel (\ref{eq:si_kernel}) with $(A,B,C,D)$ being a realization of $G^0(q)$, which is
a $2$nd order DT system having two stable real poles, i.e.,
\begin{align}\label{eq:highordcanonical}
G^0(q) = \frac{\bar bq}{(q+a_1)(q+a_2)},
\end{align} where $\bar b\in\R$ and $|a_i|<1,i=1,2$.
We consider the inverse of the kernel matrix $K$ defined by $K_{i,j}=k^\text{SI}(i,j)$ with $i,j=1,\dots,10$.  By Proposition \ref{prop:markov4si_kernel}, $K^{-1}$ should be a 2-band matrix, which is confirmed by
Fig. \ref{fig:bandedinverse}.
\end{Example}

\begin{Remark}
The Markov property of a kernel and its
associated special structure (the banded inverse of the kernel
matrix) can be used to develop numerically more stable and
efficient implementations for this kernel based regularization method, see e.g., \cite[Section 5]{CCL17}.
\end{Remark}

\begin{Remark}
The CT case is not discussed here because according to \cite[p.
217]{RasmussenW:06}, regular sampling of a CT Gaussian Markov
process entropy in general would not lead to a DT Gaussian Markov
process. That is to say, even if a  CT SI kernel with Markov
property is constructed, its corresponding kernel matrix evaluated
on the sampling instants may not have banded matrix inverse.
\end{Remark}

\section{A Case Study: Impulse Response with Damped
Oscillation}\label{sec:casestudy}

In this section, we consider the estimation of impulse responses
with damped oscillation and demonstrate how to design kernels from
the proposed two perspectives.

\subsection{The Machine Learning Perspective}

As shown in Section \ref{sec:ml_perspective}, the machine learning
perspective treats the impulse response as a function, and designs
AMLS kernels with the rank-1 kernel and the stationary kernel to
account for the decay and varying rate of the impulse response,
respectively. Now we show that by further exploiting the rank-1
kernel or the stationary kernel, we can design AMLS kernels capable
to embed the extra prior knowledge that the impulse response has
damped oscillation.

\begin{figure}[!t]
\begin{center}
\includegraphics[width=3.4in, height=3in]{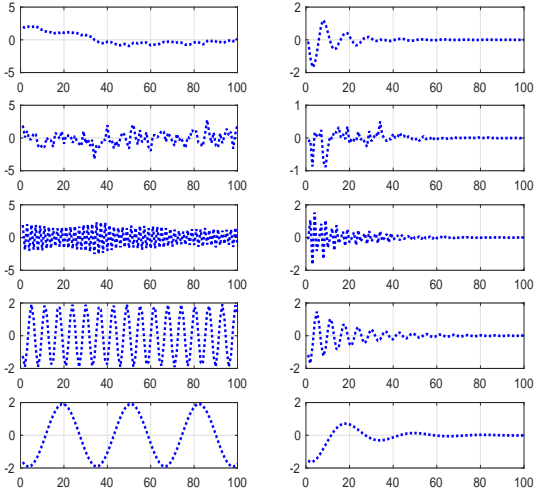}
\end{center}
\caption{Realizations of DT zero mean Gaussian process with AMLS
kernels \eqref{eq:2nd_amls}. For each row, the left panel shows the
realizations of the DT zero mean Gaussian process with the IS kernel
$k^{c}(t-s)$ as covariance function and the right panel shows the
realization of the DT zero mean Gaussian process with the AMLS
kernel as covariance function. The top three rows correspond to the
AMLS kernel (\ref{eq:2nd_amls_oscdecay}) with $c=1$,
$\lambda=0.9^{\frac{1}{2}}$, $\omega=0.2\pi$ and
$\rho=0.99,0,-0.99$, respectively. The bottom two rows correspond to
the AMLS kernel (\ref{eq:2nd_amls_oscstat}) with $c=1$,
$\lambda=0.9^{\frac{1}{2}}$ and $\alpha=\pi,0.2\pi$, respectively.
}\label{fig:exm2}
\end{figure}

To have oscillation behavior in the AMLS kernel, we can choose to
have it either in the stationary kernel or in the rank-1 kernel,
i.e., $b(t)$. For example, we can choose $b(t)$ as an exponential
decay function, i.e., $b(t)=c\lambda^t$ and then choose
\eqref{eq:ctkernel_dampedoscillation_special} as the stationary
kernel, because \eqref{eq:ctkernel_dampedoscillation_special} has
oscillation behavior. Or we can choose the stationary kernel in
\eqref{eq:DC_decomp} and $b(t)$ as an exponential decay function
with oscillation, i.e.,
\begin{align}\label{eq:oscdecay_k1}
  b(t) = c^{\frac{1}{2}}\lambda^t[\cos(\omega t)+1+\epsilon],
\end{align} where $c>0$, $0\leq \lambda< 1$, $\omega\geq0$, and $\epsilon>0$ is a tiny number to ensure that $b(t)>0$ for $t\in
X$. The above idea leads to the following two kernels, respectively:
\begin{subequations}\label{eq:2nd_amls}
\begin{align}
  k^{\text{AMLS-2Os}}(t,s) &= c\lambda^{t+s}\cos(\alpha |t-s|),\label{eq:2nd_amls_oscstat}\\
  k^{\text{AMLS-2Od}}(t,s) &= c\lambda^{t+s}[\cos(\omega t)+1+\epsilon]\nonumber\\&\qquad\times[\cos(\omega s)+1+\epsilon]\rho^{|t-s|}\label{eq:2nd_amls_oscdecay},
\end{align}
\end{subequations}
Fig. \ref{fig:exm2} illustrates that the AMLS kernels
\eqref{eq:2nd_amls} are capable to describe functions with damped
oscillation behavior.


\subsection{The System Theory Perspective}

As shown in Section \ref{sec:st_perspective}, the system theory
perspective associates the impulse response with an LTI system, and
designs SI kernels with the nominal model to embed the prior
knowledge on the LTI system. Now we show that by choosing the
nominal model to be a second order LTI system with a pair of complex
conjugate poles, we design a SI kernel capable to model impulse
responses with damped oscillation.

More specifically, we choose the transfer function of the nominal
model $G^0(q)$ in Fig. \ref{fig:multiuncertainty} to be
\begin{align}\label{eq:si_2order} G^0(s) &= \frac{1}{s^2 + 2w_0\xi s
+ w_0^2}
\end{align} where $\omega_0>0$ and $0\leq\xi<1$. By setting
$\alpha = w_0\xi$ and $\beta = w_0\sqrt{1-\xi^2}$, a state space
model for (\ref{eq:si_2order}) is described by $(A,B,C,D)$ with
\begin{align*}
  A=\left[
      \begin{array}{cc}
        0 & 1 \\
        -\alpha^2-\beta^2 & -2\alpha \\
      \end{array}
    \right], B=\left[
               \begin{array}{c}
                 0 \\
                 1 \\
               \end{array}
             \right], C=\left[
                        \begin{array}{cc}
                          1 & 0 \\
                        \end{array}
                      \right],D=0.
\end{align*} Finally, setting $Q=I_2$ and $b(t)=e^{-\gamma t}$ with $\gamma>0$ yields that the SI kernel (\ref{eq:si_kernel_ct}) takes the form: 
\begin{align}\label{eq:2nd_sikernel}
\nonumber  k^{\text{SI-2Od}}&(t,s) = e^{-\alpha(t+s)}[\cos(\beta
t)\cos(\beta s)\\\nonumber&+\frac{\alpha}{\beta}\sin(\beta
(t+s))+\frac{\alpha^2+1}{\beta^2}\sin(\beta t)\sin(\beta s)]
\nonumber\\\nonumber&+\frac{e^{-\alpha(t+s)}\cos(\beta(t-s))(e^{2(\alpha-\gamma)\min\{t,s\}}-1)}{4\beta^2(\alpha-\gamma)}\\\nonumber
&+\frac{e^{-\alpha(t+s)}}{2\beta^2\sqrt{4\beta^2+4(\alpha-\gamma)^2}}\times
[\cos(\phi+\beta(t+s))-\\&e^{2(\alpha-\gamma)\min\{t,s\}}\cos(2\beta\min\{t,s\}-\phi-\beta(t+s))],\nonumber\\&
\phi=\arccos\left(\frac{2(\alpha-\gamma)}{4\beta^2+4(\alpha-\gamma)^2}\right).
\end{align}

\subsection{Numerical Simulation}

To illustrate that the AMLS kernls (\ref{eq:2nd_amls}) and the SI
kernel (\ref{eq:2nd_sikernel}) are capable to model LTI systems with
strong oscillation, we consider the following numerical example.

\subsubsection{Test Data-bank}

The way in \cite{CCLP14} is used to generate the test systems and
data bank. In particular, we first generate 200 test systems with
strong oscillation:
\begin{align}\label{eq:test_sys}
  G(q) = \frac{q+0.99}{q} \sum_{i=1}^{N_r+1} G_i(q)
\end{align} where \begin{align*}
  G_i(q) = K_i \frac{q+0.9}{(q-p_i)(q-p_i^*)},\ i=1,\dots,N_r,
\end{align*} and $G_{N_r+1}(q)$ is a 4th order system randomly generated by the function \texttt{drmodel} in MATLAB with its poles inside the disk of radius 0.95. The parameters $N_r,p_i,K_i$ are randomly generated as follows:
$N_r\sim \mathcal U[3, 8]$, $K_i\sim \mathcal U[2, 10]$,
$p_i=\rho_ie^{j[\phi_0+\frac{\pi}{2N_r}(i-1)]}$ with $\phi_0\sim
\mathcal U[0, \frac{\pi}{2}]$ and $\rho_i\sim \mathcal U[0.8,
0.99]$.

For each of the 200 DT systems \eqref{eq:test_sys}, we generated the
test data as follows. The function \texttt{idinput} in MATLAB is
used to generate a random Gaussian input $u(t)$ with normalized band
$[0,\ 0.95]$ and length 210. The chosen DT system is simulated with
$u(t)$ to get the noise-free output $G(q)u(t)$, and $G(q)u(t)$ is
then perturbed by a white Gaussian noise, whose standard deviation
is equal to $\mathcal U[0.5, 1]$ times the standard deviation of
$G(q)u(t)$.

In this way, we generate 200 test DT systems and for each system a
data set with 210 data points.

\subsubsection{Simulation Results}

The following model fit is used to measure how good the estimated
impulse response is:
\begin{align*} \emph{fit} =100\left(1 -
\left[\frac{\sum^{100}_{k=1}|g_k^0-\hat{g}_k|^2
}{\sum^{100}_{k=1}|g_k^0-\bar{g}^0|^2}\right]^{\frac{1}{2}}\right),\
\bar{g}^0=\frac{1}{100}\sum^{100}_{k=1}g^0_k
\end{align*} where $g_k^0$ and $\hat{g}_k$ are the true impulse response and its estimate at the $k$th time instant, respectively.

The AMLS kernels (\ref{eq:2nd_amls}) and the SI kernel
(\ref{eq:2nd_sikernel}) are compared with the TC kernel
(\ref{eq:TC}), the DC kernel (\ref{eq:DC}) and the SS kernel
(\ref{eq:SS}) enriched with a second order parametric part proposed
in \cite{PCN11} and denoted by \text{SSp} below.

\begin{figure}[!t]
\begin{center}
\includegraphics[width=3.3in, height=2in]{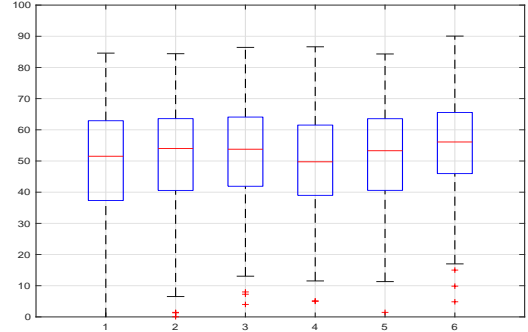}
\end{center}
\caption{Boxplot of the model fits for the TC, DC, SSp, AMLS-2Os,
AMLS-2Od and SI-2Od kernels [from left to right]. The six kernels
have 5, 3, 3, 3, 4, and 3 fits smaller than zero, respectively,
which are not shown in the boxplot for better display.}
\label{fig:boxplot_2rdSIkernel}
\end{figure}

The simulation result is summarized below. The average model fits
for the tested kernels are shown below:
\begin{center}\tiny
\begin{tabular}{|l|c|c|c|c|c|c|}
  \hline
   & TC & DC & SSp & AMLS-2Os & AMLS-2Od & SI-2Od \\\hline
  Avg. Fit & 47.5 & 50.0 & 50.9 & 48.4 & 49.7 & 53.3 \\
  \hline
\end{tabular}
\end{center}
The distribution of the model fits are shown in Fig.
\ref{fig:boxplot_2rdSIkernel}.

\subsubsection{Findings}

For the test systems and data bank, we have the following findings.
The SI kernel (\ref{eq:2nd_sikernel}) is best among all the tested
kernels in both the average accuracy and robustness and thus is best
to model systems with strong oscillation. The AMLS kernel
(\ref{eq:2nd_amls_oscdecay}) works well as it is better than the TC
kernel in both the average accuracy and robustness and it is just a
bit worse than the DC kernel in the average accuracy. The AMLS
kernel (\ref{eq:2nd_amls_oscstat}) works not very well, though it
has better average accuracy than the TC kernel. One possible
explanation why the SI kernel (\ref{eq:2nd_sikernel}) is best to
model impulse response with strong oscillation is that the prior
knowledge embedded in the SI kernel (\ref{eq:2nd_sikernel}) is
closest to the truth. This may be true because it makes use of the
prior knowledge from a system theory perspective directly by
choosing the nominal model to be a second order system with a pair
of complex conjugate poles.

\begin{Remark}
To embed the prior knowledge such as multiple distinct time
constants and multiple resonant frequencies,  we proposed to use the
multiple kernel introduced in \cite{CALCP14} where the multiple
kernel is a conic combination of some fixed kernels. The fixed
kernels can be instances of the SS, TC, DC and SI kernels
(\ref{eq:si_kernel}) evaluated on a grid in their hyper-parameter
space. The interested readers are referred to  \cite{CALCP14} for
more details.
\end{Remark}

\section{Conclusion}\label{sec:conclusion}

Kernel methods or regularization methods to estimate the impulse
response of a linear time invariant systems is a most useful
relatively recent addition to the techniques of system
identification. Earlier results have demonstrated that important
improvements in estimation quality can be achieved with kernels
devised in a more or less ad hoc way. In this contribution, we have
focused on systematic mechanisms and design concepts for how to
construct kernels that are capable of adjusting its hyperparameters
to capture the unknown system's properties in useful ways. The two
main avenues to such thinking have been a \emph{machine learning
perspective} focusing on function properties of the impulse response
and a \emph{system theory perspective} focusing on the LTI system
that produces the impulse response. This has lead to understanding
and insights into general properties of the earlier suggested
methods, e.g. that the so called DC and SS kernels (derived from
more ad hoc thinking) belong to the class of amplitude modulated
locally stationary kernels (Proposition \ref{prop:ss&dc_amls}) and
that they are simulation induced from certain LTI systems
(Proposition \ref{prop:si_kernel4ssdc}). They are also related to
maximum entropy optimal choices (Proposition
\ref{thm:maxent4si_kernel}) which is a valuable feature.

The take-home message of this contribution is as follows. The issue
of kernel design should relate to the type of the prior knowledge
and different types of prior knowledge should lead to different ways
to design the corresponding kernels. Here, a machine learning
perspective and a system theory perspective are introduced
accordingly:
\begin{itemize}
\item Machine Learning perspective: If the impulse response is
treated as a function and the prior knowledge is on its decay and
varying rate, then we can design the amplitude modulated locally
stationary (AMLS) kernel. In particular, we design a rank-1 kernel
and a stationary kernel to account for the decay and varying rate of
the impulse response, respectively. Moreover, by further exploiting
the rank-1 kernel or the stationary kernel, it is possible to design
AMLS kernels capable to embed more general prior knowledge.

\vspace{1mm}

\item System Theory perspective: If the impulse is associated with an
LTI system and the prior knowledge is that the LTI system is stable
and may be overdamped, underdamped, have multiple distinct time
constants and resonant frequencies and etc., then we can design the
simulation induced (SI) kernel. In particular, the nominal model is
used to embed the prior knowledge, the uncertainty is assumed to be
stable and finally, the multiplicative uncertainty configuration is
used to take into account both the nominal model and the
uncertainty, and is simulated with an impulsive input to get the SI
kernel or equivalently the zero-mean Gaussian process to model the
impulse response.

\end{itemize}

Finally, finding a suitable kernel structure is only one leg of the
kernel-based regularization method. Tuning its hyperparameters
regardless of structure is the other main topic. This has been
discussed in some detail e.g. in \cite{PDCDL14} and \cite{PC16}.
Some related new asymptotic results are recently presented in
\cite{MCL16}.

%
%
%
%

\begin{ack}
The authors would like to thank Prof. Lennart Ljung for his suggestions and comments at the early stage of this project. The authors also would like to thank Prof. Gianluigi Pillonetto, Prof. Alessandro
Chiuso, Prof. H{\aa}kan Hjalmarsson, Dr. John Latarie and Dr. Giulio Bottegal for
their helpful comments. 
\end{ack}

\bibliographystyle{unsrt}
\bibliography{ref}                                                      

\appendix

\renewcommand{\thesection}{A}

\section*{Proof of Proposition \ref{prop:AMLS}}

\emph{Part (a)}.  For any $n\in\mathbb N$ and for any $t_i\in X$, $i=1,\dots,n$, let $\bar b$ be the column vector containing $b(t_1),\dots,b(t_n)$ in order. Then the kernel matrix $K^{d}$ defined by $K^{d}_{i,j}=k^{d}(t_i,t_j)$ is rewritten as $K^{d}=\bar b\bar b^T$. Clearly, $K^{d}$ is positive semidefinite, and moreover $\rank (K^{d})=1$. So $k^{d}(t,s)$ is a rank-1 kernel by definition.
The identity (\ref{eq:rank-1kernel_identity}), \emph{Part (b) and Part (c)} can be verified in a straightforward way.  
%

\section*{Proof of Proposition \ref{prop:ss&dc_amls}}

We give the proof for the CT case with $X=\{t|t\geq0\}$. The results for the DT case with $X=\mathbb N$ hold by noting that the DT kernel is the CT kernel restricted to $\mathbb N$.

From (\ref{eq:SS_decomp}) and (\ref{eq:DC_decomp}), we see the proof will be done if it can be shown that the kernels $k^{c}(t-s)$ in (\ref{eq:SS_decomp}) and (\ref{eq:DC_decomp}) are stationary kernels. To show this, note from e.g. \cite[page 85]{RasmussenW:06}
that $e^{-\beta |t-s|}$ with $\beta>0$ is called the exponential
kernel and belongs to the class of Mat\'{e}rn covariance functions
with order $1/2$. So it remains to show $k^{c}(t-s)$
in (\ref{eq:SS_decomp}) is also a kernel. Note that the exponential kernel is an isotropic stationary (IS) kernel (see Section \ref{sec:ISkernel} for its definition). Below we show that $k^{c}(t-s)$
in (\ref{eq:SS_decomp}) is actually also an IS kernel. As shown in
\cite{Genton02}, the spectral representation of an IS kernel
$k^{c}(t,s)$ takes the form of
\begin{align}\label{eq:isotropic} 
k^{c}(t,s) &= \int_0^{\infty}
\cos(\omega |t-s|) F(d\omega)
\end{align} where $F$ is any nondecreasing bounded function. 
For the
exponential kernel $e^{-\beta |t-s|}$ with $\beta>0$,
(\ref{eq:isotropic}) is satisfied with the spectral density
\begin{align*}
  f(\omega) = \frac{\beta}{\pi(\beta^2+\omega^2)}.
\end{align*} So $k^{c}(t-s)$ in (\ref{eq:SS_decomp}) with $\lambda=\exp(-\beta/2)$ satisfies
(\ref{eq:isotropic}) with  a well defined spectral density described
by \begin{align*}
 \frac{3}{2} &\frac{\frac{1}{2}\beta}{\pi((\frac{\beta}{2})^2+\omega^2)} -\frac{1}{2}
 \frac{\frac{3}{2}\beta}{\pi((\frac{3\beta}{2})^2+\omega^2)}\\& = \frac{3\beta}{4\pi}
 \left(\frac{1}{(\frac{\beta}{2})^2+\omega^2} -
 \frac{1}{(\frac{3\beta}{2})^2+\omega^2}\right) \geq 0, \quad
 \forall \omega
\end{align*} Therefore, $k^{c}(t-s)$ in (\ref{eq:SS_decomp}) is
an IS kernel. This completes the proof.

\section*{Proof of Proposition \ref{thm:amls_stability}}

We give the proof for the CT case with $X=\{t|t\geq0\}$. The proof for the DT case with $X=\mathbb N$ can be derived in a similar way.

\emph{Part (a).} Assume that $b(t) \in \mathcal L_1$, i.e., $\int_0^\infty b(t)dt<\infty$. Since $|k^{c}(t-s)|\leq 1$ for any $t,s\geq0$. Then from
$$\int_{0}^\infty \left|\int_{0}^\infty k(t,s)dt\right|ds \leq \int_{0}^\infty b(s) ds\int_{0}^\infty b(t)dt<\infty,$$
and by Corollary \ref{coro:stability}, the AMLS kernel (\ref{eq:AMLS}) is stable.

\emph{Part (b).} Assume that the AMLS kernel (\ref{eq:AMLS}) is stable, i.e., $\mathcal H_k\subset \mathcal L_1$. We need a lemma to prove the result.
\begin{Lemma}\emph{\cite[p. 16]{EPP00}, \cite[p. 37]{CS02}}\label{lem:simplifiedmercer} Assume that $\bar k(t,s)$ with $t,s\in X$ is a positive semidefinite kernel and moreover, there exists a sequence of positive numbers $\nu_i$ and linearly independent functions $\psi_i(t)$
 such that $\bar k(t,s) = \sum_{i=1}^\infty \nu_i \psi_i(t)\psi_i(s)$, where the convergence is absolute and uniform on $Y_1\times Y_2$ with $Y_1,Y_2$ being any compact subsets of $X$. Then \begin{align*}
   \mathcal H_{\bar k} = \left\{ g| g=\sum_{i=1}^\infty \mu_i\psi_i, \sum_{i=1}^\infty \frac{\mu_i^2}{\nu_i}<\infty\right\}
 \end{align*} and for any $f_1,f_2\in\mathcal H_{\bar k} $ with $f_1=\sum_{i=1}^\infty c_i\psi_i$ and $f_2=\sum_{i=1}^\infty d_i\psi_i$, the inner product over $\mathcal H_{\bar k}$ is defined as
  $$\langle f,g\rangle_{\mathcal H_{\bar k}} = \sum_{i=1}^\infty \frac{c_id_i}{\nu_i}.$$
\end{Lemma}


Noting (\ref{eq:k2_mercer}) and by Lemma \ref{lem:simplifiedmercer}, we have \begin{align}\label{eq:Hk2}
\mathcal H_{k^{c}} = \left\{ h| h=\sum_{i=1}^\infty \mu_i\phi_i, \sum_{i=1}^\infty \frac{\mu_i^2}{\lambda_i}<\infty\right\}.
\end{align}
Then from (\ref{eq:k2_mercer}), the AMLS kernel (\ref{eq:AMLS}) is rewritten as
\begin{align}
k(t,s)=\sum_{i=1}^\infty \lambda_i\rho_i(t)\rho_i(s), \rho_i(t) = b(t)\phi_i(t),
\end{align}  and by Lemma \ref{lem:simplifiedmercer}, we have \begin{align}\label{eq:Hk}
\mathcal H_{k} = \left\{ g| g=\sum_{i=1}^\infty \mu_i\rho_i, \sum_{i=1}^\infty \frac{\mu_i^2}{\lambda_i}<\infty\right\}.
\end{align}
By (\ref{eq:Hk}) and (\ref{eq:Hk2}), we have \begin{align}\label{eq:Hk_alter}
  \mathcal H_k = \{g| g(t)=b(t)h(t), t\geq0, h\in\mathcal H_{k^{c}}\}.
\end{align} Since $\mathcal H_k\subset \mathcal L_1$, for any $g\in \mathcal H_k$, $g\in\mathcal L_1$, i.e.,
\begin{align}\label{eq:proof_int1}\int_{0}^\infty |g(t)|dt=\int_{0}^\infty b(t)|h(t)|dt <\infty,\end{align}
where the equality follows from (\ref{eq:Hk_alter}) and $h\in \mathcal H_{k^{c}}$ is the function associated with $g\in \mathcal H_k$.


Since $k^c$ is not stable, $\mathcal H_{k^{c}}\nsubseteq \mathcal L^1$ and there exists $h\in\mathcal H_{k^{c}}$ but $h\notin\mathcal L^1$. For such $h$, if there exits an $\epsilon>0$ such that $b(t)\geq\epsilon$ for all $t\geq0$, then \begin{align*}
  \int_{0}^\infty |g(t)|dt=\int_{0}^\infty b(t)|h(t)|dt \geq \epsilon \int_{0}^\infty |h(t)|dt = \infty,
\end{align*} which contradicts with (\ref{eq:proof_int1}). This completes the proof.

\section*{Proof of Proposition \ref{prop:amls_sikernel}}


To prove the result, we need a lemma.
\begin{Lemma}\label{lem:ssr_dt} \emph{\cite{Astrom:70,GL00}} Consider a zero mean stationary Gaussian process $h(t)$ with covariance function $k^{c}(t-s)$. Assume that $k^{c}(t-s)$ has rational power spectral density $\Psi(\omega)$\footnote{Check the statement of Theorem \ref{prop:amls_sikernel} for the definition of rational power spectral density. }. Then the following results hold:
\begin{itemize}
 \item For the DT case, there exists a rational function $G$ which has all poles inside the unit circle and all zeros inside or on the unit circle such that
\begin{align} \Psi(\omega) = G(e^{i\omega})G(e^{-i\omega})
\end{align} where $|\cdot|$ denotes the modulus of a complex number. Moreover,
let the quintuple $(\bar A,\bar B,\bar C,\bar D,\bar Q)$ be a state space realization of
$G(e^{i\omega})$, i.e., $G(e^{i\omega})=\bar C(e^{i\omega}
I_{\dim(\bar A)} - \bar A)^{-1}\bar B+\bar D$, where $\dim(\bar
A)$ is the dimension of $\bar A$ and $I_{\dim(\bar A)}$ is the
identity matrix with dimension $\dim(\bar A)$. Then the stationary Gaussian
process $h(t)$ has the following state space representation
\begin{subequations}\label{eq:stationary_dt}
\begin{align}\label{eq:SS_state_dt} x(t+1) &= \bar Ax(t)+ \bar Bw(t),
                                                              x(0)\sim\N(0,\bar Q)
                                                              \\
                                                              h(t)&=\bar Cx(t)+ \bar Dw(t)
\end{align} where $w(t)$ is the zero mean white Gaussian noise with unit variance and $\bar Q$ is the solution of the Lyapunov equation
\begin{align}\label{eq:lyapunov_eq_dt}\bar Q=\bar A\bar Q\bar A^T+\bar B\bar B^T
\end{align}
\end{subequations}

\item For the CT case, there exists a rational function $G$ which has all poles in the
left half plane and all zeros in the left half plane or on the
imaginary axis such that
\begin{align} \Psi(\omega) = G(i\omega)G(-i\omega).
\end{align}
Moreover, let the quintuple $(\bar A,\bar B,\bar C,\bar D,\bar Q)$ be a state space realization of
$G(i\omega)$, i.e., $G(i\omega)=\bar C(i\omega I_{\dim{\bar A}} - \bar A)^{-1}\bar B+\bar D$.
Then the stationary Gaussian
process $h(t)$ has the following state space representation
\begin{subequations}\label{eq:stationary}
\begin{align}\label{eq:SS_state} \dot x(t) &= \bar Ax(t)+ \bar Bw(t),
                                                              x(0)\sim\N(0,\bar Q)
                                                              \\
                                                              h(t)&=\bar Cx(t)+\bar Dw(t)
\end{align} where $w(t)$ is the zero mean white Gaussian noise with unit power spectral density and $\bar Q$ is the solution of the Lyapunov equation
\begin{align}\label{eq:lyapunov_eq}\bar A\bar Q+\bar Q\bar A^T+\bar B\bar B^T=0
\end{align}
\end{subequations}

\end{itemize}
\end{Lemma}

\emph{\textbf{Proof:} For the DT case, the first part is a result of the Spectral Factorization Theorem
\cite[Theorem 3.1 in Chapter 4]{Astrom:70} and the second part is a result of the
Representation Theorem \cite[Theorem 3.2  in Chapter 4]{Astrom:70} and
\cite[Theorem 5.3 and Equation (5.92)]{GL00}. For the CT case, the first part is a result of the Spectral Factorization Theorem
\cite[Theorem 5.1  in Chapter 4]{Astrom:70} and the second part is a result of the
Representation Theorem \cite[Theorem 5.2  in Chapter 4]{Astrom:70} and
\cite[Theorem 5.3]{GL00}.\hfill $\lozenge$.}

For the DT case, let $z(t)=c^{\frac{1}{2}}\lambda^tx(t)$. Then simple calculation shows that the AMLS kernel (\ref{eq:AMLS}) with $k^{d}(t,s)=c\lambda^{t+s}$ is in the form of (\ref{eq:si_kernel}) with $A=\lambda\bar A$, $B=\lambda\bar B$, $C=\bar C$, $D=\bar D$, $Q=c\bar Q$ and $b(t)=c^{\frac{1}{2}}\lambda^t$. For the CT case, let $z(t)=c^{\frac{1}{2}}\e^{-\beta t}x(t)$ with $\lambda=e^{-\beta}$. Then simple calculation shows that the AMLS kernel (\ref{eq:AMLS}) with $k^{d}(t,s)=c\lambda^{t+s}$ is in the form of (\ref{eq:si_kernel}) with $A=\bar A-\beta I_{\dim{\bar A}}$, $B=\bar B$, $C=\bar C$, $D=\bar D$, $Q=c\bar Q$ and $b(t)=c^{\frac{1}{2}}\e^{-\beta t}$.

\section*{Proof of Proposition \ref{prop:si_kernel4ssdc}}

We only sketch the proof for the SS kernel and the proof for the DC kernel can be derived in a similar way.

For the DT case,
the IS kernel $k^{c}(t-s)$ in (\ref{eq:SS_decomp}) has the power spectral
density
\begin{align*} &\Psi(\omega) =
\sum_{\tau=-\infty}^{+\infty} k^c(\tau) e^{-i\omega \tau}\\&=
\frac{3}{2} \frac{1-\lambda^2}{(1-\lambda e^{-i\omega})(1-\lambda
e^{i\omega})} - \frac{1}{2} \frac{1-\lambda^6}{(1-\lambda^3
e^{-i\omega})(1-\lambda^3 e^{i\omega})}\\&=\frac{(1-\lambda^2)^3}{2}
\frac{2+2\lambda^2+\lambda(e^{-i\omega}+e^{-i\omega})}{(1-\lambda
e^{-i\omega})(1-\lambda e^{i\omega})(1-\lambda^3
e^{-i\omega})(1-\lambda^3 e^{i\omega})}\\&=
\frac{(1-\lambda^2)^3}{2}
\frac{(\bar ae^{-i\omega}+\bar b)(\bar ae^{i\omega}+\bar b)}{(1-\lambda
e^{-i\omega})(1-\lambda e^{i\omega})(1-\lambda^3
e^{-i\omega})(1-\lambda^3 e^{i\omega})}
\end{align*}

By spectral factorization technique, we have $\Psi(\omega) =
G(e^{i\omega})\overline{G(e^{i\omega})}$ with \begin{align*}
G(e^{i\omega})= \sqrt{\frac{(1-\lambda^2)^3}{2}}
\frac{(\bar ae^{-i\omega}+\bar b)}{(1-\lambda e^{-i\omega})(1-\lambda^3
e^{-i\omega})}.
\end{align*}

For the CT case, the IS kernel $k^{c}(t-s)$ has the
power spectral density
\begin{align*}\nonumber
\Psi(\omega) 
&=\int_{-\infty}^{+\infty}
\left(\frac{3}{2}e^{-\frac{1}{2}\beta|\tau|} -
\frac{1}{2}e^{-\frac{3}{2}\beta|\tau|}\right) e^{-iw\tau}
d\tau\nonumber\\
&=\left|\frac{3^{\frac{1}{2}}\beta^{\frac{3}{2}}}{(i\omega +
\frac{1}{2}\beta)(i\omega + \frac{3}{2}\beta)}\right|^2\triangleq
|G(i\omega)|^2.
\end{align*} From the
realization theory of linear systems, see e.g., \cite{Chen99}, we can derive the corresponding state-space model representation of the IS kernel $k^{c}(t-s)$, based on which and by using the argument in the end of the proof of Proposition \ref{prop:amls_sikernel}, we derive (\ref{eq:SS_ssrform_dt}) and (\ref{eq:SS_ssrform_ct}), respectively.

\section*{Proof of Proposition \ref{prop:si_kernel_stability_suf}}

For the DT case, note that there exists $l>0$ such that \begin{align*}
&|CA^tQ(A^{s})^TC^T|\leq l|\bar\lambda|^{t+s},\\
&|D^2
b(t)b(s)\delta(t-s)|\leq l|\bar\lambda|^{\frac{t+s}{2}},\\&|Db(t)\sum_{k=0}^{s-1}\delta(t-k)b(k)B^T(A^{s-1-k})^TC^T|\leq l|\bar\lambda|^{\frac{t+s}{2}},\\&\left|\sum_{k=0}^{\min\{t,s\}-1} b(k)^2
CA^{t-1-k}BB^T(A^{s-1-k})^TC^T\right|\\&\qquad\leq l(|\bar\lambda|^{t+s}+|\bar\lambda|^{t+s-\min\{t,s\}}).
\end{align*} Then
it is easy to show
$\sum_{t=1}^\infty \sum_{s=1}^\infty$ $\left|k^{\text{SI}}(t,s)\right|
<\infty$, and thus the SI kernel (\ref{eq:si_kernel_dt}) is stable by Corollary \ref{coro:stability}.

For the CT case, note that there exists $l>0$ such that \begin{align*}
&|Ce^{At}Q(e^{As})^TC^T|\leq le^{-|\text{Re}(\bar \lambda)|(t+s)},\\
&\left|C\int_0^{\min\{t,s\}} b^2(\tau)
e^{A(t-\tau)} B B^T (e^{A(s-\tau)})^T
d\tau C^T\right|\\&\qquad\leq l(e^{-|\text{Re}(\bar \lambda)|(t+s)}+e^{-|\text{Re}(\bar \lambda)|(t+s-\min\{t,s\})}).
\end{align*} Then
it is easy to show
 $\int_0^\infty \int_0^\infty \left|k^{\text{SI}}(t,s)\right|$ $dtds
<\infty$, and thus the SI kernel (\ref{eq:si_kernel_ct}) is stable by Corollary \ref{coro:stability}.

\section*{Proof of Proposition \ref{thm:maxent4si_kernel}}

We only give the proof for the case $D\neq0$ and the proof for the case
(b) is similar and thus omitted.

By chain rule, the differential entropy in
(\ref{eq:maxent4ssm_Dnonzero}) becomes
\begin{align}\label{eq:maxent_proof1}
&H(\bar z(0),\bar g(0),\bar g(1),\cdots,\bar g(s))=H(\bar z(0))+
H(\bar g(0)|\bar z(0))\nonumber
\\& + \sum_{t=1}^{s} H(\bar g(t)|\bar z(0),\bar g(0),\cdots,\bar g(t-1))
\end{align}
Note from (\ref{eq:maxent4ssm_intmf_nonzero}) that
\begin{align*}
H(\bar g(0)|\bar z(0))&=H(f(0)Db(0)+C\bar z(0)|\bar z(0))\\
&=H(f(0)Db(0)|\bar z(0))\\
&=H(f(0)|\bar z(0)) + \log |Db(0)|
\end{align*} where the first equation follows because translation does not change the differential entropy
and the second equation follows because of the scaling property of
differential entropy. Analogously, we have
\begin{align*}
H(\bar g(t)&|\bar z(0),\bar g(0),\cdots,\bar g(t-1))\\&=H(\bar g(t)|\bar z(0),f(0),\cdots,f(t-1))\\
&=H(f(t)Db(t)|\bar z(0),f(0),\cdots,f(t-1))\\
&=H(f(t)|\bar z(0),f(0),\cdots,f(t-1)) + \log |Db(t)|
\end{align*}
Therefore, (\ref{eq:maxent_proof1}) is rewritten as
\begin{align*}
&H(\bar z(0),\bar g(0),\bar g(1),\cdots,\bar g(s))=H(\bar z(0))+
H(f(0)|\bar z(0))\nonumber
\\& + \sum_{t=1}^{s} H(f(t)|\bar z(0),f(0),\cdots,f(t-1))+\sum_{t=0}^{s} \log
|D b(t)|\nonumber\\
&=H(\bar z(0),f(0),f(1),\cdots,f(s))+\sum_{t=0}^{s} \log
|D b(t)|
\end{align*} Since the second term in the above equation is independent of $\bar z(0)$ and
$f(t)$, $t=0,\cdots,s$, the MaxEnt problem
(\ref{eq:maxent4ssm_Dnonzero}) is equivalently rewritten as
\begin{subequations}\label{eq:maxent4ssm_Dnonzero2}
\begin{align}
&\maximize_{\bar z(0),f(t)} \quad H(\bar z(0),f(0),f(1),\cdots,f(s))\\
&\mbox{subject to}\nonumber  \\
&\quad \mean{\bar z(0)}=0,\mean{f(t)}=0, \nonumber\\&\quad \cov{\bar z(0)} = Q, \var{f(t)}= 1, t =
0,\cdots,s.
\end{align}
\end{subequations}
Note that the constraints in (\ref{eq:maxent4ssm_Dnonzero2}) are
separable with respect to $\bar z(0),f(t)$, $t=0,\cdots,s$, and
that $H(\bar z(0),f(0),f(1),\cdots,f(s))$ is maximized if $\bar
z(0),f(0)$, $f(1),\cdots,f(s)$ are independent with each other.
Therefore, (\ref{eq:maxent4ssm_Dnonzero2}) is equivalent to
\begin{align*}
&\maximize_{\bar z(0)} \quad H(\bar z(0))\\
&\mbox{subject to}\  \mean{\bar z(0)}=0, \cov{\bar z(0)} = Q
\end{align*} and for $t=0,\cdots,s$,
\begin{align*}
&\maximize_{f(t)} H(f(t))\\
&\mbox{subject to}\ \mean{f(t)}=0,  \var{f(t)}= 1
\end{align*}
It is well known that multivariate normal distribution maximizes the
differential entropy over all distributions with the same
covariance, see e.g., \cite[Theorem 8.6.5]{CT12}. Then we have the
optimal solution to (\ref{eq:maxent4ssm_Dnonzero2}) is that $\bar
z(0)\sim\N(0,Q)$ and $f(t)\sim\N(0,1)$, $t=0,\cdots,s$, and
moreover, $\bar z(0),f(0),f(1),\cdots,f(s)$ are independent with
each other. Finally, (\ref{eq:maxent4ssm_intmf_nonzero}) implies
$\bar g(t) = CA^t \bar z(0) + \sum_{k=0}^{t-1}CA^{t-1-k}B
b(k)f(k) +
  Db(t)f(t)
$, same as $g(t)$ in (\ref{eq:si_kernel_dt}). This completes the proof.

\section*{Proof of Corollary \ref{coro:maxent4si_dc}}

From Proposition \ref{prop:si_kernel4ssdc}, the DC kernel is a SI kernel with
$A,B,C,D,Q,b(t)$ given in (\ref{eq:DC_ssrform_dt}), which implies that Theorem \ref{thm:maxent4si_kernel} holds for the DC kernel. Comparing (\ref{eq:maxent4ssm_intmf_nonzero}) with (\ref{eq:maxent4dc}) shows that the proof is completed if we can show
\begin{align}\nonumber
&  \cov{\bar z(0)} = Q\quad  \Longrightarrow\quad \var{\bar g(0)}=c,\\
&  \var{f(t)}= 1 \Longrightarrow  \var{\bar g({t})-\lambda^{1/2}\rho \bar
g({t-1})}\nonumber\\&\qquad\qquad\qquad\qquad
 = c(1-\rho^2)\lambda^{t}, t=1,\cdots, s\label{eq:proof_dcmaxent_1}
\end{align} The first line of (\ref{eq:proof_dcmaxent_1}) is straightforward. In what follows, we consider the second line of (\ref{eq:proof_dcmaxent_1}). Clearly, the results holds for the case either $\lambda=0$ or $\rho=0$. So we further consider below the case where  $\lambda\neq0$ and $\rho\neq0$. It is easy to see that for $t=0,\dots,s-1$,  \begin{align*}
  \bar g(t+1) &= ACA^t\bar z(0) + A\sum_{k=0}^{t-1}CA^{t-1-k}Bb(k)f(k)\\&\quad+ACA^{-1}Bb(t)f(t)+Db(t+1)f(t+1)\\
  &=ACA^t\bar z(0) + A\sum_{k=0}^{t-1}CA^{t-1-k}Bb(k)f(k)\\&\quad+ADb(t)f(t)+Db(t+1)f(t+1)\\
  &=A\bar g(t)+Db(t+1)f(t+1)
\end{align*} where the second equality holds because $CA^{-1}B=D$. Therefore (\ref{eq:proof_dcmaxent_1}) holds, which completes the proof.

\section*{Proof of Proposition \ref{prop:dc_markov}}

We need a lemma to prove this result.
\begin{Lemma}\label{lemma:markovproperty}
Consider the $p$th order Gaussian Markov process (\ref{eq:gmp}).
Then the following results hold:
\begin{enumerate}
\item[(a)] For any $t\in\mathbb N$ and $k>p$,
$x(t)$ and $x(t+k)$ are conditionally independent given $x(s)$ with
$s\neq t$, $s\neq t+k$ and $s\in\mathbb N$.

\item[(b)] Let $i_1< i_2< \cdots< i_n$ with $n>p+1$ be any strictly increasing subsequence taken from $\mathbb N$. Then the covariance matrix of $[x(i_1)\ \cdots\ x(i_n)]^T$  is a $p$-band matrix.
\end{enumerate}

\end{Lemma}

\emph{\textbf{Proof:} Part (a) follows from the $p$th order Markov property of $x(t)$ and
the fact that $w(t)$ is white Gaussian noise. To prove Part (b), recall from e.g., \cite{Dempster72} that, for Gaussian random variables, the zeros in the inverse of the
covariance matrix indicate conditional independence of the two
corresponding elements conditioned on the remaining ones. To be
specific, consider a Gaussian random variable with mean $m$ and
covariance matrix $K$:
\begin{align}\label{eq:gaussian4markov}
\left[
  \begin{array}{ccc}
    x &
    y &
    z
  \end{array}
\right]^T\sim\N(m,{K}),
\end{align} where $x,y\in\R$, $z\in\R^n$ and ${K}\in\R^{(n+2)\times
(n+2)}$. 
Then for a given $z$, $x,y$ are conditionally independent if and only if $({K}^{-1})_{1,2}=({K}^{-1})_{2,1}=0$, where $(K^{-1})_{i,j}$ denote the $(i,j)$th element of $K^{-1}$.
Therefore, Part (b) follows from the above observation and
part (a).
\hfill $\lozenge$}

Note from Proposition \ref{prop:si_kernel4ssdc} that $g(t)$ can be written in the following form \begin{align*}
 &z(t+1)= \lambda^{\frac{1}{2}}\rho z(t)+\lambda^{\frac{1}{2}}c^{\frac{1}{2}}\lambda^{\frac{t}{2}}w(t),z(0)\sim\N(0,\frac{c}{1-\rho^2})\nonumber\\&g(t)=\rho(1-\rho^2)^{\frac{1}{2}}z(t) + (1-\rho^2)^{\frac{1}{2}}c^{\frac{1}{2}}\lambda^{\frac{t}{2}}w(t).
\end{align*} which implies that (\ref{eq:dc_gmp}) holds. Clearly, (\ref{eq:dc_gmp}) is in the form of (\ref{eq:gmp_1}) with $p=1$ and thus $g(t)$ is a Markov process with
order 1.  Finally, the result that the kernel matrix of the DC kernel has a 1-band matrix inverse follows from Lemma \ref{lemma:markovproperty}.

\section*{Proof of Proposition \ref{prop:markov4si_kernel}}

By using the realization of $G^0(q)$ in controllable canonical form, the DT SI kernel (\ref{eq:si_kernel}) is written as follows:
\begin{align*}  z(t+1)&= \left[
                                                      \begin{array}{cccccc}
                                                        -a_1\  \cdots\  -a_{n-1} & -a_n\\
                                                        I_n & {\bf
                                                        0}_{n\times1}
                                                      \end{array}
                                                    \right]
 z(t)\\&\qquad\nonumber\qquad +
\left[
  \begin{array}{c}
    1 \\
     {\bf
                                                        0}_{n\times1} \\
  \end{array}
\right]
 b(t) w(t) \\ g(t)&=\left[
                                           \begin{array}{ccccc}
                                             \bar b & 0 & 0 & \cdots & 0\\
                                           \end{array}
                                         \right]
 z(t),z(0)\sim\N(0,Q).
\end{align*}

From the above equation, we have
\begin{align}\label{eq:proof4markov_int1} g(t+1) &= -\bar b\sum_{i=1}^n
a_iz_i(t) + \bar b b(t)w(t)
\end{align}
where $z_i(t)$ is the $i$th element of $z(t)$. Note that
\begin{align*}
  \bar bz_{i}(t) = g(t-i+1), i=1,\cdots,n
\end{align*} Therefore, (\ref{eq:proof4markov_int1}) becomes
\begin{align*}
g(t+1)= -\sum_{i=1}^n  a_ig(t-i+1) + \bar b b(t)w(t)
\end{align*}
which is in the form of (\ref{eq:gmp_2}) with order $n$. Thus the Gaussian process $g(t)$ in (\ref{eq:si_kernel}) is a
Markov process with order $n$ and the fact that the kernel has an $n$-band
matrix inverse follows from Lemma  \ref{lemma:markovproperty}.

\end{document}